\documentclass[12pt,twoside]{article}

\def\TheAuthor{Patrick Bruno}                                   
\def\TheTitle{Berry phase effects in magnetism}                        
\def\TheHeading{Berry phase effects in magnetism}                                      
\def\TheInstitute{Max-Planck-Institut f\"{u}r Mikrostrukturphysik}       
\def\TheAddress{Weinberg 2, D-06120 Halle, Germany}              
\def\ThePart{E9}                                               
\def\TheChapter{}                                            

\usepackage{ferienschule_04}                                  

\usepackage{bm}

\newcommand{\para}{\hspace*{0cm}}

\begin{document}

\MakeTitel           

\vspace*{-4\baselineskip}

Lecture notes published in ``\emph{Magnetism goes nano}'', Lecture
Manuscripts of the 36th Spring School of the Institute of Solid
State Research, edited by Stefan~Bl\"{u}gel, Thomas~Br\"{u}ckel,
and Claus M.~Schneider (Forschungszentrum J\"{u}lich, 2005).

\vspace*{4\baselineskip}

\tableofcontents     


\newpage


\section{Introduction}\label{sec:intro}

\para In 1983, Berry made the surprising discovery that a quantum
system adiabatically transported round a closed circuit
$\mathcal{C}$ in the space of external parameters acquires,
besides the familiar dynamical phase, a non-integrable phase
depending only on the geometry of the circuit $\mathcal{C}$
\cite{Berry1984}. This Berry phase, which had been overlooked for
more than half a century, provides us a very deep insight on the
geometric structure of quantum mechanics and gives rise to various
observable effects. The concept of the Berry has now become a
central unifying concept in quantum mechanics, with applications
in fields ranging from chemistry to condensed matter physics
\cite{Shapere1989, Bohm2003}.

The aim of the present lecture is to give an elementary
introduction to the Berry phase, and to discuss its various
implications in the field of magnetism, where it plays an
increasingly important role. The reader is referred to specialized
textbooks \cite{Shapere1989, Bohm2003} for a more comprehensive
presentation of this topic.

\section{Parallel transport in geometry}\label{sec:geo}

The importance of the Berry phase stems from the fact that it
reveals the intimate geometrical structure underlying quantum
mechanics. It is therefore appropriate to start with an
introduction of the fundamental concept of \emph{parallel
transport} in a purely geometrical context; we follow here the
discussion given by Berry in Ref.~\cite{Berry1990}.

This is best illustrated by means of a simple example. Consider a
surface $\Sigma$ (e.g., a plane, a sphere, a cone, etc.) and a
vector constrained to lie everywhere in the plane tangent to the
surface. Next, we wish to transport the vector on the surface,
\emph{without rotating it around the axis normal to the surface},
as illustrated in Fig.~1. We are interested, in particular in the
case, in which the arrow is transported round a closed circuit
$\mathcal{C}\equiv (1\rightarrow 2 \rightarrow 3 \rightarrow 1)$.
We may encounter two different situations: (i) if the surface is
flat, as on Fig.~1(a), then the arrow always remains parallel to
its original orientation, and therefore is unchanged after
completion of the circuit $\mathcal{C}$; (ii) if, however, the
surface $\Sigma$ is curved as on Fig.~1(b,c), the arrow, being
constrained to lie in the local tangent plane, cannot remain
parallel to its original orientation, and after completion of the
circuit $\mathcal{C}$, it is clearly seen to have been rotated by
an angle $\theta (\mathcal{C})$, a phenomenon referred to as
\emph{anholonomy}.

\begin{figure}[h]\label{fig:para_transp}
\begin{center}
\epsfig{file=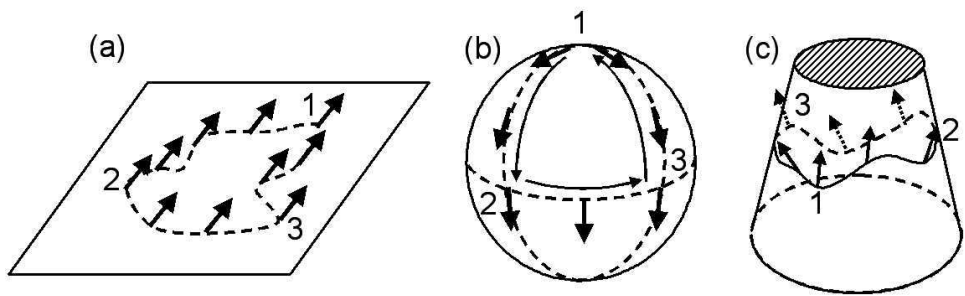,width=0.8\textwidth} \caption{Sketch of
parallel transport on (a) a plane, (b) a sphere, and (c) a cone.}
\end{center}
\end{figure}

Let us now formalize this procedure. The arrow is represented by a
tangent unit vector $\mathbf{e}^1$, transported along a circuit
$\mathcal{C} \equiv \{\mathbf{r}(t) | t=0 \rightarrow T \}$ on the
surface. Defining $\mathbf{n}(\mathbf{r})$ as the unit vector
normal to the surface at point $\mathbf{r}$, we define a second
tangent unit vector $\mathbf{e}^2 \equiv \mathbf{n \times e}^1$,
which is also parallel transported on the surface along
$\mathcal{C}$. The 3 unit vectors
$(\mathbf{n},\mathbf{e}^1,\mathbf{e}^2)$ form an orthonormal
reference frame. As $\mathbf{e}^1$ and $\mathbf{e}^2$ are
transported, they have to rotate with an angular velocity
$\bm{\omega}$ (to be determined) if the surface is not flat, i.e.,
the equation of motion of $\mathbf{e}^1$ and $\mathbf{e}^2$ is
\begin{equation}
\dot{\mathbf{e}^r}= \bm{\omega}\times \mathbf{e}^r\ \ \ (r=1,2),
\end{equation}
where the overdot indicates the time derivative. One can easily
see that in order to fulfill the requirements that $\mathbf{e}^1$
and $\mathbf{e}^2$ remain tangent unit vectors (i.e.,
$\mathbf{e}^r \cdot \mathbf{n}=0$, $(r=1,2)$) and never rotate
around $\mathbf{n}$ (i.e., $\bm{\omega}\cdot \mathbf{n}=0$), the
angular velocity has to be given by
\begin{equation}
\bm{\omega}= \mathbf{n \times \dot{n}}.
\end{equation}
The law of parallel transport is therefore
\begin{equation}\label{eq:para_transp_0}
\mathbf{\dot{e}}^r = (\mathbf{n\times \dot{n}})\times \mathbf{e}^r
= -(\mathbf{e}^r \cdot \mathbf{\dot{n}}) \mathbf{n}.
\end{equation}
This law can be expressed in a form more suitable for
generalization to the case of quantum mechanics, by defining the
complex unit vector
\begin{equation}
\bm{\phi}\equiv \frac{\mathbf{e}^1 +
\mathrm{i}\mathbf{e}^2}{\sqrt{2}} ,
\end{equation}
with
\begin{equation}
\bm{\phi}^\star \cdot \bm{\phi} =1 .
\end{equation}
The law of parallel transport now reads
\begin{equation}\label{eq:para_transp}
\bm{\phi}^\star \cdot \dot{\bm{\phi}}=0 .
\end{equation}
In order to express the rotation of the unit vectors
$(\mathbf{e}^1,\mathbf{e}^2)$ as they move around $\mathcal{C}$,
we need to chose a \emph{fixed} local orthonormal frame
$(\mathbf{n}(\mathbf{r}),\mathbf{t}^1(\mathbf{r}),\mathbf{t}^2(\mathbf{r}))$on
the surface. The normal unit vector $\mathbf{n}(\mathbf{r})$ is of
course uniquely determined by the surface, but we have an infinity
of possible choices for $\mathbf{t}^1(\mathbf{r})$ (we simply
impose that it is a smooth function of $\mathbf{r}$), which
corresponds to a gauge freedom; once we have made a choice for
$\mathbf{t}^1(\mathbf{r})$, then $\mathbf{t}^2(\mathbf{r})$ is of
course uniquely determined. We next define the complex unit vector
\begin{equation}
\mathbf{u}(\mathbf{r})\equiv \frac{\mathbf{t}^1(\mathbf{r}) +
\mathrm{i}\mathbf{t}^2(\mathbf{r})}{\sqrt{2}} ,
\end{equation}
with, of course,
\begin{equation}
\mathbf{u}^\star (\mathbf{r})\cdot \mathbf{u}(\mathbf{r}) =1 .
\end{equation}
The relation between the parallel transported frame and the fixed
one is expressed as
\begin{equation}
\bm{\phi}(t)=\exp [ -\mathrm{i}\theta(t)]\ \mathbf{u} \left(
\mathbf{r}(t)\right) ,
\end{equation}
where $\theta(t)$ is the angle by which
$(\mathbf{t}^1,\mathbf{t}^2)$ must be rotated to coincide with
$(\mathbf{e}^1,\mathbf{e}^2)$. We obtain the equation satisfied by
$\theta(t)$ by inserting the above definition in the equation of
parallel transport (\ref{eq:para_transp}), and obtain
\begin{equation}
0=\bm{\phi}^\star \cdot \dot{\bm{\phi}} = -\mathrm{i}\
\dot{\theta}\ \mathbf{u}^\star \cdot \mathbf{u} + \mathbf{u}^\star
\cdot \dot{\mathbf{u}}  .
\end{equation}
Since $\mathbf{u}^\star \cdot\mathbf{u}=1$ and then
$\mathbf{u}^\star \cdot\mathbf{\dot{u}}$ is imaginary, we get
\begin{equation}\label{eq:para_transp_1}
\dot{\theta} = \mathrm{Im} (\mathbf{u}^\star \cdot
\dot{\mathbf{u}}) ,
\end{equation}
so that
\begin{eqnarray}
\theta(\mathcal{C}) &=& \mathrm{Im} \oint_\mathcal{C}
\mathbf{u}^\star \cdot \mathrm{d}\mathbf{u}  \\
&=& -\oint_\mathcal{C} \mathbf{t}^2\cdot\mathrm{d}\mathbf{t}^1 .
\end{eqnarray}
If we choose a coordinate system $(X_1,X_2)$ on our surface
$\Sigma$, and define the vector field $\mathbf{A}(\mathbf{r})$
(usually called a \emph{connection}) on $\Sigma$ as
\begin{equation}
A_i(\mathbf{X}) \equiv \mathrm{Im}\left[ u_j^\star(\mathbf{X})
\frac{\partial u_j (\mathbf{X})}{\partial X_i} \right],
\end{equation}
where we have used Einstein'convention of summation over repeated
indices, we get
\begin{equation} \label{eq:holo_1}
\theta(\mathcal{C}) =  \oint_\mathcal{C} \mathbf{A}(\mathbf{X})
\cdot \mathrm{d}\mathbf{X} ,
\end{equation}
which constitutes the \emph{1-form} expression of the anholonomy
angle $\theta(\mathcal{C})$. The connection
$\mathbf{A}(\mathbf{X})$ depends on our particular gauge choice
for $\mathbf{t}^1(\mathbf{X})$: if we make a new choice
${\mathbf{t}^1}^\prime(\mathbf{X})$ which is brought in
coincidence with $\mathbf{t}^1(\mathbf{X})$ by a rotation of angle
$\mu (\mathbf{X})$, i.e., if we make the gauge transformation
\begin{equation}
\mathbf{u}(\mathbf{X}) \rightarrow \mathbf{u}^\prime(\mathbf{X})
\equiv \exp\left(-\mathrm{i}\mu (\mathbf{X})\right)
\mathbf{u}(\mathbf{X})  ,
\end{equation}
we obtain a new connection
\begin{equation}
A^\prime_i (\mathbf{X})\equiv \mathrm{Im}\left[
u_j\prime^\star(\mathbf{X}) \frac{\partial u_j\prime
(\mathbf{X})}{\partial X_i} \right] = A_i(\mathbf{X}) -
\frac{\partial \mu (\mathbf{X})}{\partial X_i} .
\end{equation}
However, since
\begin{equation}
\oint_\mathcal{C} \nabla \mu(\mathbf{r}) \cdot
\mathrm{d}\mathbf{r} = \oint_\mathcal{C} \mathrm{d}\mu
(\mathbf{r})=0 ,
\end{equation}
we can see that the expression (\ref{eq:holo_1}) for the
anholonomy angle $\theta(\mathcal{C})$ is indeed gauge invariant,
as it should.

\begin{figure}[h]\label{fig:stokes}
\begin{center}
\epsfig{file=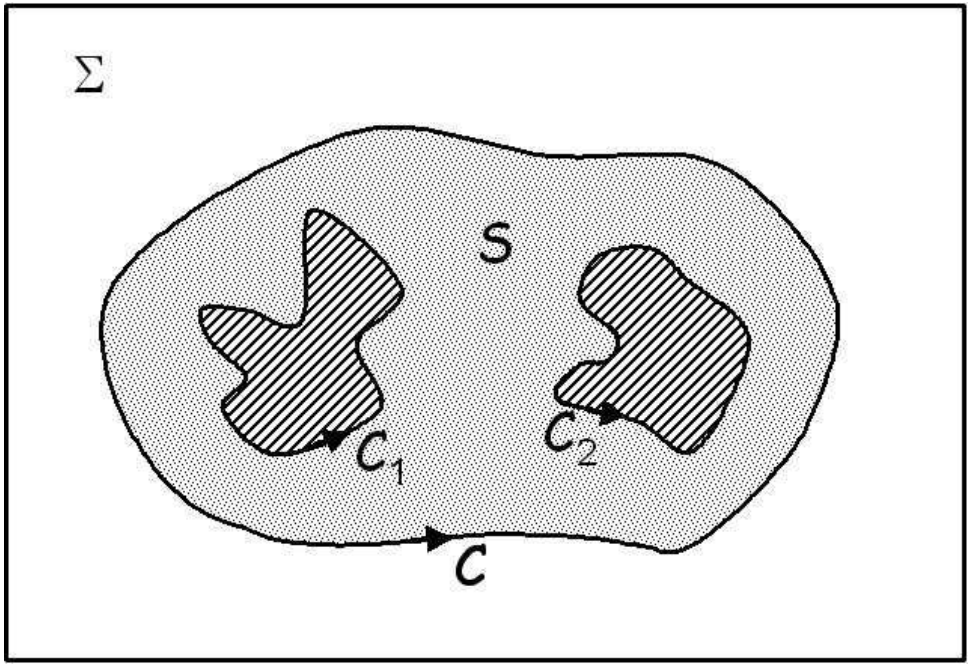,width=0.8\textwidth} \caption{Sketch of a
non simply connected surface $\Sigma$, with 2 holes (hatched
areas), limited by the contours $\mathcal{C}_1$ and
$\mathcal{C}_2$.}
\end{center}
\end{figure}

\para
A more intuitive understanding of the anholonomy angle may be
obtained if we use Stokes' theorem to express it as a surface
integral. In doing so, however, we should pay attention to the
possible existence of holes in the surface $\Sigma$. If this is
the case, $\Sigma$ is said to be non simply connected. An example
is sketched on Fig.~2, where the surface $\Sigma$ has 2 holes
limited by the contours $\mathcal{C}_1$ and $\mathcal{C}_2$
(hatched areas on Fig.~2). Applying Stokes theorem, we then obtain
\begin{equation} \label{eq:holo_2}
\theta(\mathcal{C}) = \int \!\!\! \int_\mathcal{S} B(\mathbf{X})
\mathrm{d}X_1\,\mathrm{d}X_2  + \sum_i N_i (\mathcal{C})
\theta(\mathcal{C}_i) .
\end{equation}
where the surface $\mathcal{S}$ is the subset of the surface
$\Sigma$ limited by the circuit (dotted area on Fig.~2),
$\mathcal{C}$, $N_i (\mathcal{C}$ is the winding number of circuit
$\mathcal{C}$ around the hole $i$ (i.e., the difference between
the number of turns in counterclockwise and clockwise directions),
\begin{equation}
\theta(\mathcal{C}_i) \equiv \oint_{\mathcal{C}_i}
\mathbf{A}(\mathbf{X}) \cdot \mathrm{d}\mathbf{X}
\end{equation}
is the anholonomy angle of circuit $\mathcal{C}_i$ and
\begin{eqnarray}
B(\mathbf{X})&\equiv& \left( \frac{\partial A_2}{\partial X_1} -
\frac{\partial A_1}{\partial X_2} \right) \nonumber \\
&=& \mathrm{Im} \left[ \frac{\partial \mathbf{u}^\star}{\partial
X_1}\cdot \frac{\partial \mathbf{u}}{\partial X_2} -
\frac{\partial \mathbf{u}^\star}{\partial X_2} \cdot\frac{\partial
\mathbf{u}}{\partial X_1}\right] .
\end{eqnarray}
Equation (\ref{eq:holo_2}) constitutes the \emph{2-form}
expression of the anholonomy angle $\theta(\mathcal{C})$. One can
see immediately that, unlike the connection
$\mathbf{A}(\mathbf{X})$, the quantity $B(\mathbf{X})$ is gauge
invariant. The geometrical meaning of $B(\mathbf{X})$ stems from
its relation to the \emph{Gaussian curvature} $K$ of $\Sigma$ at
point $\mathbf{X}$, i.e.,
\begin{equation}
B(\mathbf{X})\mathrm{d}X_1 \mathrm{d}X_2 = K\mathrm{d}S\equiv
\frac{\mathrm{d}S}{R_1 (\mathbf{X})\, R_2 (\mathbf{X})} ,
\end{equation}
where $R_1(\mathbf{X})$ and $R_2(\mathbf{X})$ are the principal
curvature radii at point $\mathbf{X}$. In the case of the sphere,
this is easily checked by explicit calculation, taking the usual
spherical angles $(\theta , \varphi)$ as variables $(X_1,X_2)$.
Since the Gaussian curvature is related to the solid angle
$\Omega$ of spanned by the normal unit vector $\mathbf{n}$ by
\begin{equation}
B = \frac{\mathrm{d}^2 \Omega}{\mathrm{d}X_1\, \mathrm{d}X_2}
\end{equation}
we finally get
\begin{equation} \label{eq:holo_3}
\theta(\mathcal{C}) - \sum_i N_i (\mathcal{C})
\theta(\mathcal{C}_i) = \int \!\!\! \int_\mathcal{S}
\frac{\mathrm{d}^2 \Omega}{\mathrm{d}X_1\, \mathrm{d}X_2}
\mathrm{d}X_1\,\mathrm{d}X_2 = \int \!\!\! \int_\mathcal{S}
\mathrm{d}^2 \Omega = \Omega (\mathcal{S}),
\end{equation}
where $\Omega (\mathcal{S})$ is the solid angle described by the
normal vector $\mathbf{n}$ on the surface $\mathcal{S}$. That the
above results hold not only for a sphere, but for any surface can
be understood easily from the following argument:
Eq.~(\ref{eq:para_transp_0}) shows that the trajectory of the
parallel transported tangent vectors is entirely determined by the
trajectory of the normal unit vector $\mathbf{n}$ along
$\mathcal{C}$. We can therefore map the trajectory $\mathcal{C}$
on the surface $\Sigma$ to a trajectory $\mathcal{C}^\prime$ on
the sphere of unit radius $S^2$, by mapping each point of $\Sigma$
onto the point of $S^2$ with the same normal vector $\mathbf{n}$.
This implies that we can restrict ourselves to studying the case
of parallel transport on $S^2$ and obtain conclusions valid for
parallel transport on any surface $\Sigma$.

Let us examine these results for the examples sketched on Fig.~1.
For the case of the plane, the anholonomy of course trivially
vanishes. For the sphere, the anholonomy angle is given by the
solid angle $\Omega (\mathcal{S})$, and is therefore a
\emph{geometric} property of the circuit $\mathcal{C}$; this can
be easily checked by making the following experiment: take your
pen in you left hand, and raise your arm above you head, the pen
pointing in front of you; then rotate your arm until it is
horizontal in front of you, without twisting your hand; then
rotate it by 90$^\mathrm{o}$ to your left; finally rotate your arm
back to the vertical  (pay attention to never twist your hand in
whole process); the pen is now pointing to your left, i.e., it has
rotated by $4\pi/8 =\pi /2$. For the case of the cone, the
Gaussian curvature vanishes everywhere (a cone can be fabricated
by rolling a sheet of paper), so that the the anholonomy angle is
in fact a \textit{topological} property of the circuit
$\mathcal{C}$, being given by the winding number of the circuit
$\mathcal{C}$ around the cone (multiplied by the solid angle of
the cone).

\section{Parallel transport in classical mechanics: Foucault's
pendulum and the gyroscope}\label{sec:classical}

Let us now consider the famous experiment of Foucault's pendulum
that demonstrated the earth's rotation. If the pendulum trajectory
is originally planar (swinging oscillation), the vertical
component of the angular momentum vanishes. Since forces exerted
on the pendulum (gravity and wire tension) produce a vanishing
vertical torque, the vertical component of the angular momentum
has to be conserved. The absence of any vertical torque imposes
that the swing plane has to follow a law of parallel transport as
the direction of gravity slowly changes due to the earth's
rotation. Therefore, within one day it rotates by an angle equal
to the solid angle described by the vertical $2\pi (1- \cos\theta
)$, where $\theta$ is the colatitude.

The parallel transport may also affect the phase of the periodic
motion of the Foucault pendulum or the rotation phase of a
gyroscope, but also the phase of their periodic motion. Let us
consider a gyroscope whose rotation axis is constrained to remain
parallel to the axis $\mathbf{n}$; let us now move the rotation
axis $\mathbf{n}$ round a closed circuit $\mathcal{C}$. The
rotation angle of the gyroscope will be the sum of the
\emph{dynamic rotation angle} $\omega t$ and of a \emph{geometric
anholonomy angle} $\theta (\mathcal{C})$ equal to the solid angle
described by the rotation axis. Thus if we have two synchronous
gyroscopes and perform different circuits with the rotation axes,
they will eventually be dephased with respect to each other, an
effect that could easily be observed by stroboscopy. This
geometric anholonomy angle is known as Hannay's angle
\cite{Hannay1985, Berry1985}. If the Foucault pendulum is given a
conical oscillation, instead of a planar swing, then we have
exactly the same situation as described above for the gyroscope,
and the rotation angle will have an anholonomy excess angle given
by the solid angle described by the vertical. Thus, two identical
Foucault pendula (i.e., of same length) with circular oscillations
in opposite directions will have slightly different oscillation
frequencies, and will progressively get dephased with respect to
each other. The swinging motion of the usual Foucault may be view
as the superposition of circular motions in opposite direction, so
that the rotation of the swinging plane may be viewed as resulting
from the above mentioned frequency shift.

\section{Parallel transport in quantum mechanics: the Berry
phase}\label{sec:quantum}

\para
Let us now consider a quantum mechanical system described by a
Hamiltonian controlled by a set of external parameters $(R_1, R_2,
\ldots )$, which we describe collectively as a vector $\mathbf{R}$
in some abstract parameter space. Physically, the external
parameters may be magnetic or electric fields, etc. For each value
$\mathbf{R}$ of the external parameters, the Hamiltonian
$H(\mathbf{R})$ has eigenvalues $E_n(\mathbf{R})$ and eigenvectors
$\left| n(\mathbf{R})\right\rangle$ satisfying the independent
Schr\"{o}dinger equation, i.e.,
\begin{equation}
H(\mathbf{R}) \left| n(\mathbf{R})\right\rangle = E_n(\mathbf{R})
\left| n(\mathbf{R})\right .
\end{equation}
The eigenvectors $\left| n(\mathbf{R})\right\rangle$ are defined
up to an arbitrary phase, and there is \emph{a priori} no
particular phase relation between eigenstates corresponding to
different values of the parameter $\mathbf{R}$. We make a
particular choice for the phase of the eigenstates, simply
requiring that $\left| n(\mathbf{R})\right\rangle$ varies smoothly
with $\mathbf{R}$ in the region of interest. It may happen that
the eigenstates we have chosen are not single valued functions of
$\mathbf{R}$. If this happens, special care must be given to this
point.

\para
Let us perform an adiabatic closed circuit $\mathcal{C} \equiv
\left\{ \mathbf{R}(t) | t =0 \rightarrow T \right\}$ in the
parameter space. The adiabatic theorem \cite{Messiah1991} tells us
that if the rate of variation of the external parameters is low
enough, a system that is initially in the $n$th stationary state
$|n\rangle$ (assumed non-degenerate) of the Hamiltonian will
remain continuously in the state $|n\rangle$. The condition of
adiabaticity is that the stationary state under consideration
remains non-degenerate, and the rate of variation of the
Hamiltonian is low enough to make the probability of transition to
another state $|m\rangle$ vanishingly small, i.e.,
\begin{equation}
\hbar|\langle m|\dot{H}|n\rangle| \ll |E_m - E_n|^2 \ \ \ \forall
\, m \neq n .
\end{equation}
Then of course, if one performs a closed adiabatic circuit
$\mathcal{C}$, the system has to return to its original state.

\para
Berry \cite{Berry1984} asked the following question: what will be
the phase of the state after completion of the circuit
$\mathcal{C}$ ? It may be difficult at first sight to realize that
this question may be of any interest. Indeed, the expectation
value of any observable quantity $A$,
\begin{equation}
\langle A\rangle \equiv \langle \psi |A| \psi\rangle
\end{equation}
does not depend of the phase of $|\psi \rangle$. This lack of
interest is certainly the main reason why the Berry phase was
(almost \footnote{Some early precursor work on effects related to
the Berry phase include notably Pancharatnam's work on optical
polarization \cite{Pancharatnam1956}, Aharonov and Bohm's work on
the phase due to the electromagnetic potential vector
\cite{Aharonov1959}, and Mead and Truhlar's work on the molecular
Aharonov-Bohm effect in the Born-Oppenheimer theory of molecular
vibrations \cite{Mead1979}.}) completely overlooked for more than
half a century of quantum mechanics.

\para
So, following Berry, taking
\begin{equation}
|\psi (t=0) \rangle\equiv \left| n(\mathbf{R}(t=0))\right\rangle
\end{equation}
we express the state $|\psi (t)\rangle$ at a latter time $t$ as
\begin{equation}
|\psi (t)\rangle \equiv \exp \left[ \frac{-\mathrm{i}}{\hbar}
\int_0^t \mathrm{d}t^\prime \ E_n(\mathbf{r}(t^\prime )) \right]
|\phi_n (t) \rangle ,
\end{equation}
i.e., we introduce an auxiliary wavefunction $|\phi_n (t) \rangle$
with a zero dynamical phase. Using the time-dependent
Schr\"{o}dinger equation,
\begin{equation}
\mathrm{i}\hbar |\dot{\psi} (t)\rangle = H(t) |\psi (t)\rangle ,
\end{equation}
and projecting it on $\langle \psi (t)|$, we get
\begin{eqnarray} \label{eq:quant_para_transp}
0&=& \langle \psi (t) | \left( H(t) - \mathrm{i}\hbar
\frac{\partial}{\partial t} \right) |\psi (t)\rangle  \nonumber \\
&=& \langle \phi_n (t) | \dot{\phi}_n (t)\rangle ,
\end{eqnarray}
where we have used the relation
\begin{equation}
\langle \psi (t) | H(t) |\psi (t)\rangle = E_n(t) ,
\end{equation}
which follows from the adiabatic theorem. Equation
(\ref{eq:quant_para_transp}) shows that the wavefunction $|\phi_n
(t) \rangle$ obeys a quantum mechanical analogue of the law of
parallel transport (\ref{eq:para_transp}).

\para
In complete analogy with the problem of parallel transport on a
surface, we now express the parallel transported state $|\phi_n
(t) \rangle$ in terms of the fixed eigenstates $\left|
n(\mathbf{R})\right\rangle$ as
\begin{equation}
|\phi_n (t) \rangle \equiv \exp ((\mathrm{i}\gamma_n (t)) \left|
n(\mathbf{R})\right\rangle ,
\end{equation}
where the phase $\gamma_n (t)$ plays the same role as the angle
$-\theta (t)$ for the problem of parallel transport on a surface.
We then immediately get the equation of motion of $\gamma_n (t)$,
i.e.,
\begin{equation}
\dot{\gamma}_n (t) = \mathrm{i} \langle n| \dot{n}\rangle =
-\mathrm{Im} \langle n(\mathbf{R}(t))|
\frac{\mathrm{d}}{\mathrm{d}t}  n(\mathbf{R}(t))\rangle ,
\end{equation}
which is analogous to Eq.~(\ref{eq:para_transp_1}).

\para
Finally, the answer to the question originally asked by Berry
is
\begin{equation}
| \psi (T) \rangle = \exp \left[\mathrm{i}(\delta_n +
\gamma_n(\mathcal{C}))\right] | \psi (0) \rangle ,
\end{equation}
where
\begin{equation}
\delta_n \equiv \frac{-1}{\hbar}\int_0^T E_n(\mathbf{R}(t))\,
\mathrm{d}t
\end{equation}
is the dynamical phase, and
\begin{equation}
\gamma_n(\mathcal{C})) \equiv -\mathrm{Im}\left[ \oint_\mathcal{C}
\langle n(\mathbf{R}) | \partial_\mathbf{R} | n(\mathbf{R})
\rangle \cdot \mathrm{d}\mathbf{R} \right] - \alpha_n
(\mathcal{C})
\end{equation}
is the Berry phase. The last term in the latter equation arises
when the states $| n(\mathbf{R}) \rangle$ are not a single-valued
function of $\mathbf{R}$ in the region of interest of the
parameter space\footnote{This term was absent in Berry's original
paper \cite{Berry1984}, because the basis states $| n(\mathbf{R})
\rangle$ were assumed to be single-valued.}, and is given by
\begin{equation}
\alpha_n (\mathcal{C}) = \mathrm{i} \ln \left[ \langle
n(\mathbf{R}(0))|  n(\mathbf{R}(T)) \rangle\right] .
\end{equation}
We shall omit this term below, and consider only the case of
single-valued basis states.

We note the very close analogy between the result obtained for
quantum and classical systems. The dynamical phase of a quantum
system is analogous to the rotation angle $\omega T$ in classical
mechanics, whereas the Berry phase is analogous to Hannay's angle
(they both arise from the anholonomy of parallel transport).

\para
Defining the connection $\mathbf{A}^n(\mathbf{R})$ as
\begin{equation}
\mathbf{A}^n(\mathbf{R}) \equiv - \mathrm{Im} \left[ \langle
n(\mathbf{R}) | \partial_\mathbf{R} n(\mathbf{R}) \rangle \right]
,
\end{equation}
we re-express the Berry phase as
\begin{equation} \label{eq:Berry_1form}
\gamma_n(\mathcal{C})) \equiv \oint_\mathcal{C}
\mathbf{A}^n(\mathbf{R}) \cdot \mathrm{d}\mathbf{R} ,
\end{equation}
which constitutes the 1-form expression of the Berry phase. The
latter clearly depends only on the geometry of the circuit
$\mathcal{C}$. The connection $\mathbf{A}^n(\mathbf{R})$ is not
gauge invariant: if we make a new choice for the phase of the
reference state, i.e.,
\begin{equation}
| n(\mathbf{R}) \rangle^\prime = \exp(-i\mu(\mathbf{R}) ) |
n(\mathbf{R}) \rangle ,
\end{equation}
with a single-valued function $\mu(\mathbf{R})$, we obtain a
different connection
\begin{equation}
{\mathbf{A}^n}^\prime(\mathbf{R}) = \mathbf{A}^n(\mathbf{R}) +
\partial_\mathbf{R} \mu(\mathbf{R}) .
\end{equation}
However, the Berry phase $\gamma_n(\mathcal{C})$ is gauge
invariant, as it should.

\para
As for the geometric parallel transport on surfaces, we may obtain
a gauge invariant and more transparent expression by transforming
the above result to a surface integral by using Stokes' theorem.
Here too, we have to pay attention to the existence of holes in
the parameter space: if the parameter space is multiply connected,
and if the circuit $\mathcal{C}$ cannot be continuously deformed
to a point\footnote{A circuit that can be continuously deformed to
a point is said to be \emph{homotopic} to a point.}, we must take
into account terms associated with the winding of $\mathcal{C}$
around holes of the parameter space.

\para
The formulation of the Berry phase as a surface integral in a form
that is independent of a particular choice of coordinates of the
parameter space generally requires to use the mathematical
formalism of differential forms \cite{Bohm2003}, which is beyond
the scope of the present lecture. We can nevertheless obtain a
useful result without resorting to any advanced mathematics if we
make a suitable choice of coordinates of the parameter space. Let
us choose a surface $\mathcal{S}$ in the parameter space which is
bound by the circuit $\mathcal{C}$, and a parameterization
$(R_1,R_2)$ of the surface $\mathcal{S}$. Using Stokes' theorem we
then get
\begin{equation} \label{eq_Berry_2}
\gamma_n(\mathcal{C}) = \int \!\!\! \int_\mathcal{S}
B^n(\mathbf{R}) \mathrm{d}R_1\,\mathrm{d}R_2  + \sum_i N_i
(\mathcal{C}) \gamma_n(\mathcal{C}_i) ,
\end{equation}
where $\mathcal{C}_i$ are the circuits bounding the holes of the
parameter space and $N_i$ the corresponding winding numbers of the
circuit $\mathcal{C}$ around them, and where
\begin{eqnarray}
B^n(\mathbf{R})&\equiv& \left( \partial_{R_1} A^n_2 -
\partial_{R_2} A^n_1 \right) \nonumber \\
&=& - \mathrm{Im} \left[ \langle \partial_{R_1}  n(\mathbf{R}) |
\partial_{R_2} n(\mathbf{R}) \rangle -  \langle \partial_{R_2}  n(\mathbf{R}) |
\partial_{R_1} n(\mathbf{R}) \rangle\right]
\end{eqnarray}
is the \emph{Berry curvature}. In the case where the parameter
space is three-dimensional, then we can use the familiar language
of vector calculus, as in electrodynamics, and Stokes' theorem
yields
\begin{equation}
\gamma_n(\mathcal{C}) = \int \!\!\! \int_\mathcal{S}
\mathbf{B}^b(\mathbf{R}) \cdot \mathbf{n}\ \mathrm{d}S  + \sum_i
N_i (\mathcal{C}) \gamma_n(\mathcal{C}_i) ,
\end{equation}
\begin{eqnarray}
\mathbf{B}^n(\mathbf{R})&\equiv& \nabla\times  \mathbf{A}^n (\mathbf{R}) \nonumber \\
&=& - \mathrm{Im} \left[ \langle \nabla n(\mathbf{R}) | \times |
\nabla n(\mathbf{R}) \rangle\right] \\
&=& - \mathrm{Im} \sum_{m\neq n} \langle \nabla n(\mathbf{R}) |
m(\mathbf{R}) \rangle \times \langle m(\mathbf{R})
 |  \nabla n(\mathbf{R})\rangle .
\end{eqnarray}
Making use of the relation
\begin{equation}
\langle m | \nabla n \rangle = \frac{\langle m|\nabla H | n
\rangle}{E_n - E_m} ,
\end{equation}
one eventually get
\begin{equation}\label{eq:Berry_curv}
\mathbf{B}^n(\mathbf{R}) = - \mathrm{Im} \sum_{m\neq n}
\frac{\langle n(\mathbf{R})|\nabla H(\mathbf{R}) |
m(\mathbf{R})\rangle \times \langle m(\mathbf{R})|\nabla
H(\mathbf{R}) | n(\mathbf{R})\rangle}{\left( E_m(\mathbf{R}) -
E_n(\mathbf{R}) \right)^2} .
\end{equation}
Obviously, the Berry curvature is gauge invariant. As the notation
suggests, the Berry curvature $\mathbf{B}^n$ plays the role of a
magnetic field in the space of parameters, whose vector potential
is the Berry connection $\mathbf{A}^n$.

The energy denominator in Eq.~(\ref{eq:Berry_curv}) shows that if
the circuit $\mathcal{C}$ lies in a region of the parameter space
that is close to a point $\mathbf{R}^\star$ of two-fold degeneracy
involving the two states labelled $+$ and $-$, the corresponding
Berry connections $\mathbf{B}^+$ and $\mathbf{B}^+$ are dominated
by the term involving the denominator $(E_+ - E-)^2$ and the
contribution involving other states can be neglected. So, to first
order in $\mathbf{R}-\mathbf{R^\star}$, one has
\begin{equation}
\mathbf{B}_+ (\mathbf{R})= - \mathbf{B}_-(\mathbf{R}) = -
\mathrm{Im} \frac{\langle +(\mathbf{R})|\nabla H(\mathbf{R}^\star)
| -(\mathbf{R})\rangle \times \langle -(\mathbf{R})|\nabla
H(\mathbf{R}^\star) | +(\mathbf{R})\rangle}{\left( E_+(\mathbf{R})
- E_-(\mathbf{R}) \right)^2} .
\end{equation}
The general form of the Hamiltonian $H(\mathbf{R})$ of a two-level
system is (without loss of generality, we may take
$\mathbf{R}^\star =0$)
\begin{equation}
H(\mathbf{R}) \equiv  \frac{1}{2}\left(
\begin{array}{cc}
Z & X-\mathrm{i}Y \\
X+\mathrm{i}Y & -Z
\end{array}
\right) ,
\end{equation}
with eigenvalues
\begin{equation}
E_+(\mathbf{R}) = - E_-(\mathbf{R})=\frac{1}{2}R .
\end{equation}
This illustrates a theorem due to von Neumann and Wigner
\cite{Neumann1929}, stating that it is necessary to adjust 3
independent parameters in order to obtain a two-fold degeneracy
from an Hermitian matrix. The gradient of the Hamiltonian is
\begin{equation}
\nabla H = \frac{1}{2} \bm{\sigma},
\end{equation}
where $\bm{\sigma}$ is the vector matrix whose components are the
familiar Pauli matrices. Simple algebra then yields
\begin{equation}
\mathbf{B}_+ = - \mathbf{B}_- = - \frac{\mathbf{R}}{R^3} .
\end{equation}
The above Berry curvature $\mathbf{B}_\pm$ is the magnetic field
in parameter space generated by a Dirac magnetic monopole
\cite{Dirac1931} of strength $\mp 1/2$. Thus, the Berry phase
$\gamma_\pm (\mathcal{C})$ of a circuit $\mathcal{C}$ is given by
the flux of the monopole through the surface $\mathcal{S}$
subtended by the circuit $\mathcal{C}$, which, by Gauss' theorem,
is nothing else as $\mp\Omega (\mathcal{C})$, where $\Omega
(\mathcal{C})$ is the solid angle described by $\mathbf{R}$ along
the circuit $\mathcal{C}$.

\begin{figure}[h]
\label{fig:monopole}
\begin{center}
\epsfig{file=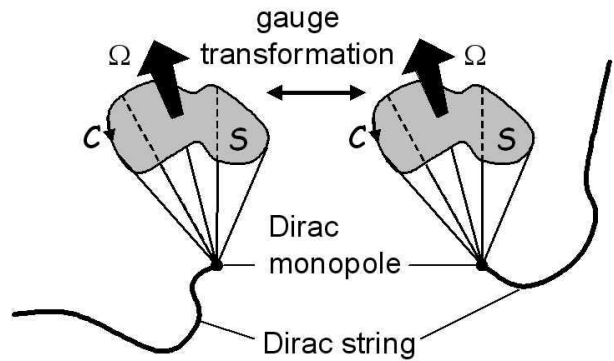,width=0.5\textwidth} \caption{Sketch showing
the flux of the Dirac monopole through the circuit $\mathcal{C}$,
and the effect of a gauge transformation.}
\end{center}
\end{figure}

The corresponding vector potential (or Berry connection)
$\mathbf{A}_\pm$ (not calculated here), has an essential
singularity along a line (Dirac string) ending at the origin, and
carrying a "flux" of magnitude $\pm 2\pi$. The position of the
Dirac string can be moved (but not removed!) by a gauge
transformation, as sketched on Fig.~3. If the Dirac string happen
to cross the cross the surface $\mathcal{S}$, the Berry phase
remains unchanged (modulo $2\pi$), so that the result is indeed
gauge invariant.

\section{Examples of Berry phase}\label{sec:examples}

\subsection{Spin in a magnetic field}

As a first example, we consider the case of a single spin (of
magnitude $S$) in a magnetic field, which is both the most
immediate application of the formal theory presented above and one
of the most frequent case encountered in experimentally relevant
situations. The Hamiltonian considered is
\begin{equation}
H(\mathbf{b})\equiv - \mathbf{b}\cdot \mathbf{S} ,
\end{equation}
with the magnetic field $\mathbf{b}$ being the external
parameters. The eigenvalues are
\begin{equation}
E_n (\mathbf{b}) = -nb ,
\end{equation}
with $2n$ integer and $-S \leq n \leq S$. For $\mathbf{b}=0$, the
$2S+1$ eigenstates are degenerate, so the circuit $\mathcal{C}$
has to avoid the origin. The Berry connection  can be calculated
using Eq.~(\ref{eq:Berry_curv}) and well known properties of the
spin operators, and one gets
\begin{equation}
\mathbf{B}^n (\mathbf{b}) = -n\frac{\mathbf{b}}{b^3} ,
\end{equation}
which is the "magnetic field" (in parameter space) of a monopole
of strength $-n$ located at the origin. The Berry phase is thus
\begin{equation}
\gamma_n (\mathcal{C}) = -n \Omega (\mathcal{C}) ,
\end{equation}
where $\Omega (\mathcal{C})$ is the solid angle described by the
field $\mathbf{b}$ along the circuit $\mathcal{C}$. For $S=1/2$,
this of course reduces to the result obtained above for the
two-level problem. Note that the Berry phase $\gamma_n
(\mathcal{C})$ depends only on the quantum number $n$ (projection
of $\mathbf{S}$ on $\mathbf{b}$) and not on the magnitude $S$ of
the spin. Note also, that while $H(\mathbf{b})$ is the most
general Hamiltonian for a spin $S=1/2$, this is not the case for a
spin $S \geq 1$; in the latter case we are restricting ourselves
here to a subspace of the full parameter space. If a more general
Hamiltonian and a wider parameter space is considered, the simple
result obtained above would not hold any more.

\subsection{Aharonov-Bohm effect}

Another example which is a great interest, both conceptually and
experimentally, is the well known Aharonov-Bohm effect
\cite{Aharonov1959}. We follow here the presentation of the
Aharonov-Bohm effect given by Berry \cite{Berry1984}.

\begin{figure}[h]
\label{fig:AB}
\begin{center}
\epsfig{file=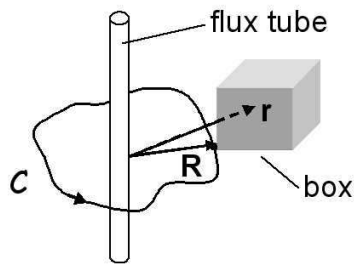,width=0.3\textwidth} \caption{Sketch
describing the Aharonov-Bohm effect.}
\end{center}
\end{figure}

Let us consider the situation depicted in Fig.~4, namely a
magnetic field confined in a tube with flux $\Phi$ and a box
located at $\mathbf{R}$ in which particles of charge $q$ are
confined. The magnetic field is vanishes everywhere outside the
flux tube, and in particular inside the box. Let
$\mathbf{A}(\mathbf{r})$ be the corresponding  vector potential.
The latter generally does not vanish in the regions of vanishing
field (unless the flux $\Phi$ is a multiple of the flux quantum
$\Phi_0 \equiv h/e$.

Let the Hamiltonian describing the particles in the box be
$H(\mathbf{p}, \mathbf{r}-\mathbf{R})$; the corresponding wave
functions, for a vanishing vector potential, are of the form
$\psi_n (\mathbf{r- R})$, with energies $E_n$ independent of
$\mathbf{R}$. When the flux is non-zero, we can chose as basis
states $| n(\mathbf{R}) \rangle$, satisfying
\begin{equation}
H(\mathbf{p} -q\mathbf{A(r)}, \mathbf{r}-\mathbf{R}) |
n(\mathbf{R}) \rangle = E_n | n(\mathbf{R}) \rangle ,
\end{equation}
whose solutions are given by
\begin{equation}
\langle \mathbf{r}|n(\mathbf{R}) \rangle = \exp \left[
\frac{\mathrm{i}q}{\hbar} \int_\mathbf{R}^\mathbf{^r}
\mathrm{d}\mathbf{r}^\prime \cdot \mathbf{A(r^\prime)}\right]
\psi_n (\mathbf{r- R}) ,
\end{equation}
where the integral is performed along a path contained in the box.
The energies $E_n$ are independent of the vector potential,
because it is always possible to find a gauge transformation that
would make it zero in the box (but not everywhere in space!).

The Hamiltonian depends on the position $\mathbf{R}$ of the box
via the vector potential. Thus our parameter space, in this
example, is nothing else than the real space, with exclusion of
the region of the flux tube. If we transport the box around a
closed circuit $\mathcal{C}$, the Berry phase will be given by
\begin{equation}
\gamma_n(\mathcal{C})) \equiv \oint_\mathcal{C}
\mathbf{A}^n(\mathbf{R}) \cdot \mathrm{d}\mathbf{R} ,
\end{equation}
with the Berry connection
\begin{eqnarray}
\mathbf{A}^n(\mathbf{R}) &\equiv& - \mathrm{Im} \left[ \langle
n(\mathbf{R}) | \partial_\mathbf{R} n(\mathbf{R}) \rangle \right]
\nonumber \\
&=& -\mathrm{Im} \int\!\!\!\int\!\!\!\int \mathrm{d}^3\mathbf{r}
\psi_n^\star (\mathbf{r-R}) \left[ \frac{-\mathrm{i}q}{\hbar}
\mathbf{A(R)} \psi_n (\mathbf{r-R}) +\partial_\mathbf{R}\psi_n
(\mathbf{r-R}) \right] \nonumber \\
&=& \frac{q}{\hbar} \mathbf{A(R)} .
\end{eqnarray}
The Berry curvature $\mathbf{B}^n(\mathbf{R})= \nabla\times
\mathbf{A}^n(\mathbf{R}) = (q/\hbar)\mathbf{B}(\mathbf{R})$ is
just given by the magnetic field, and vanishes everywhere outside
the flux tube. But because the tube region is excluded from the
allowed parameter space, the latter is multiply connected, and the
Berry phase is purely topological, being given by the winding
number $N(\mathcal{C})$ of the circuit $\mathcal{C}$ around the
flux tube, and by the flux $\Phi$:
\begin{equation}
\gamma_n(\mathcal{C}) = 2 \pi N(\mathcal{C}) \frac{q}{h}\, \Phi .
\end{equation}

The Aharonov-Bohm effect was confirmed experimentally by electron
holography by Tonomura \emph{et al.} \cite{Tonomura1986} in a
configuration where the magnetic truly vanishes, and plays an
outstanding role in the physics of mesoscopic systems, where it
gives rise to conductance oscillations and to persistent currents
in mesoscopic metallic rings threaded by a magnetic flux
\cite{Olariu1985, Aronov1987, Washburn1992} .

\section{Experimental observations of the Berry phase for a single
spin}

Let us now discuss how the Berry phase could be detected
experimentally. As already mentioned, this is not immediately
clear since the expectation value of any observable would be
independent of the phase of the system. As always when considering
phases, some kind of interference has to be observed. There
various ways in which this can be done.
\begin{itemize}
\item
Berry original proposal \cite{Berry1984} was the following: a
monoenergetic polarized beam of particles in the spin state $n$
along the magnetic field $\mathbf{b}$ is split in two beams. For
one of the beams, the field $\mathbf{b}$ is kept constant in
magnitude and direction, whereas in the second beam, the magnitude
of $\mathbf{b}$ is kept constant and its direction slowly varied
along a circuit $\mathcal{C}$ subtending a solid angle $\Omega$.
The two beams are then recombined to interfere, and the intensity
is monitored as a function of the solid angle $\Omega$. Since the
dynamical phase is the same for both beams, the phase difference
between the two beams is given purely by the Berry phase (plus a
propagation factor is determined by the phase shift for
$\Omega=0$. Although conceptually possible, it seems unlikely that
such an experiment would be feasible in practice. In particular,
it would be extremely difficult to ensure that the difference
between the dynamical phases of the two beams be smaller that the
Berry phase one wants to detect, unless some physical principle
enforces it. This kind of experiment may be said to be of type
"one state -- two Hamiltonians". This kind of experiment, being
based on interferences is truly quantum mechanical.
\item
An alternative approach, more amenable to experimental test is to
prepare the system into a superposition of two states, i.e.,
\begin{equation}
|\psi (t=0) \rangle = \alpha |n(\mathbf{R}(t=0))\rangle  + \beta
|m(\mathbf{R}(t=0))\rangle ,
\end{equation}
with $m=n-1$ and $|\alpha|^2 + |\beta|^2 =1$, for example by
polarizing it along a direction perpendicular to the field
$\mathbf{b}$. The orientation of the transverse component of the
spin is given by the angle $\theta (t=0) \equiv \arg (\beta) -
\arg (\alpha)$. The spin of course precesses at around
$\mathbf{b}$ at the Larmor frequency $\omega_L = b /\hbar$. After
completion of the circuit $\mathcal{C}$, the system state has
evolved to
\begin{equation}
|\psi (T) \rangle = \alpha \exp [\mathrm{i}(\delta_n + \gamma_n
(\mathcal{C}))] |n(\mathbf{R}(t=0))\rangle  + \beta \exp
[\mathrm{i}(\delta_m + \gamma_m (\mathcal{C}))]
|m(\mathbf{R}(t=0))\rangle ,
\end{equation}
and the polarization angle has evolved to $\theta (T) = \theta
(t=0) + \Delta\theta$ with
\begin{eqnarray}
\Delta\theta &=& \Delta\theta_\mathrm{dyn} +
\Delta\theta_\mathrm{B} , \\
\Delta\theta_\mathrm{dyn} &\equiv& \delta_m - \delta_n = \omega_L
T ,
\\
\Delta\theta_\mathrm{B} &\equiv& \gamma_m (\mathcal{C})) -
\gamma_n (\mathcal{C})) .
\end{eqnarray}
Here the angle $\Delta\theta_\mathrm{dyn}$ gives the polarization
rotation due to the Larmor precession (dynamic phase), while
$\Delta\theta_\mathrm{B}$ is the polarization rotation due to the
Berry phase accumulated along the circuit $\mathcal{C}$. Thus by
investigating how the polarization varies as the circuit
$\mathcal{C}$ is modified, the Berry phase can be detected. Such
an experiment may be said to be of the type "two states -- one
Hamiltonian". Note that this type of experiment can be interpreted
in purely classical terms \cite{Cina1986} (it bears a clear
analogy to the rotation of swinging plane of the Foucault
pendulum); this is related to the fact that only Berry phase
differences between two states, and not the absolute Berry phase
of a given state is detected.
\item
A further possibility consists in repeating the circuit
$\mathcal{C}$ in a periodic manner. Thus, the Berry phase is
accumulated linearly in time, just as the dynamical phase, and
leads to an apparent energy shift for the state $n$,
\begin{equation}
\Delta E_n = \frac{\hbar}{T} \gamma_n (\mathcal{C}) ,
\end{equation}
which gives rise to an observable shift of the transition between
to levels $n$ and $m$. Such an experiment is of type "two states
-- one Hamiltonian", too. It can also be interpreted in classical
terms and has close analogy to the period shift of a Foucault
pendulum with circular oscillation.
\end{itemize}

\subsection{Berry phase of neutrons}

The Berry phase has been observed for neutrons by Bitter and
Dubbers \cite{Bitter1987}, who used the experimental shown in
Fig.~5. A slow ($v\simeq 500\ \mathrm{m}.\mathrm{s}^{-1}$),
monochromatic, beam neutrons polarized ($P\simeq 0.97$) along an
axis perpendicular to the beam axis $z$ is injected in a cylinder
with a helical magnetic field with longitudinal component $B_z$
and transverse component $B_1$making a right-handed turn of
$2\pi$. Depending on the values of $B_z$ and $B_1$, various values
of the solid angle $\Omega$ may be achieved.

After having traversed the cylinder, the polarization of the beam
is measured, from which the Berry phase can be extracted. The
comparison of the measured Berry phase (or more precisely the
difference of Berry phase between states $S_z=+1/2$ and
$S_z=-1/2$) and of the solid angle is shown in Fig.~5. The
observation is in good agreement with the theoretical prediction.

\begin{figure}[h]
\label{fig:Bitter}
\begin{center}
\epsfig{file=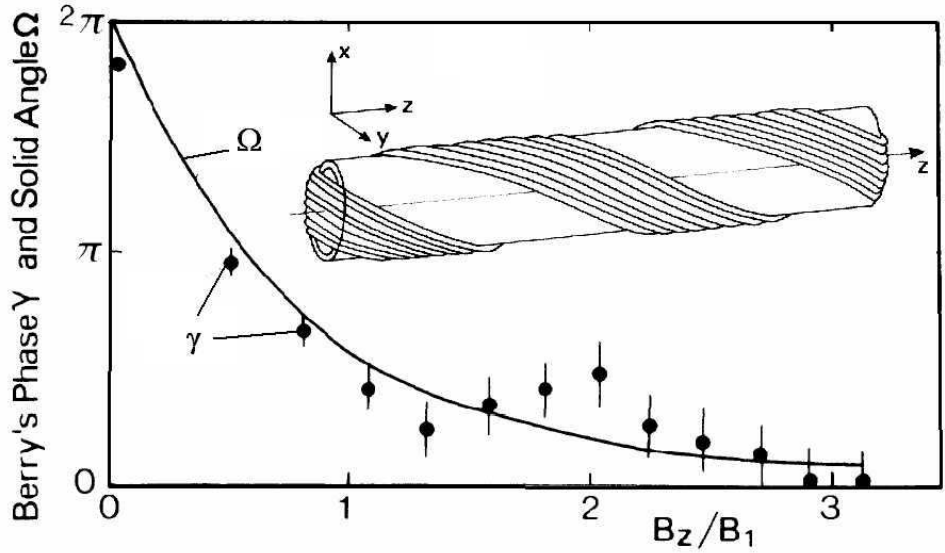,width=0.8\textwidth} \caption{Measurement of
Berry phase of neutrons. The inset shows the arrangement of the
coil giving an helical field; the neutron beam is along $z$;
length: 40~cm, diameter: 8~cm; an axial coil (not shown) produces
a field $B_z$. The curve shows the Berry phase (more precisely
$\gamma_{-1/2} - \gamma_{1/2}$) and solid angle $\Omega$ as a
function of the ratio $B_z/B_1$. From Ref.~\cite{Bitter1987}.}
\end{center}
\end{figure}

\subsection{Berry phase of photons}

The photon is a particle of spin $S=1$ and can thus experience a
Berry phase. The particularity of the photon is that, being
massless, only the states $S_z = \pm 1$ occur and that the
quantization axis is fixed by the direction of the wave vector
$\mathbf{k}$. The wavevector therefore plays the role of a
magnetic field for the photon \cite{Chiao1986}.

If the latter is constrained to make a closed circuit
$\mathcal{C}$ of solid angle $\Omega$, then a Berry phase of
$\pm\Omega$ is expected for the two circular polarizations,
respectively. If a monochromatic, linearly polarized, optical wave
\begin{equation}
|\chi\rangle = \frac{|+\rangle + |-\rangle}{\sqrt{2}} ,
\end{equation}
where $|+\rangle$ and $|-\rangle$ represent, respectively the two
circular polarization modes, is injected in a single mode optical
fiber, whose axis describes a helix of solid angle $\Omega$, then
the emerging wave will be (omitting the dynamical phase)
\begin{equation}
|\chi^\prime\rangle = \frac{\exp(\mathrm{i}\gamma_+)|+\rangle +
\exp(\mathrm{i}\gamma_-) |-\rangle}{\sqrt{2}} ,
\end{equation}
with $\gamma_+ = - \gamma_- =-\Omega$, which yields
\begin{equation}
|\langle \chi^\prime |\chi \rangle|^2 = \cos^2 \Omega .
\end{equation}
By Malus' law, this means that the polarization has rotated by an
angle $\Omega$ (the sense of rotation, when looking into the
output of the fiber is counterclockwise, i.e., dextrorotatory, for
a left-handed helix) \cite{Chiao1986}.

\begin{figure}[t]
\label{fig:Tomita}
\begin{center}
\epsfig{file=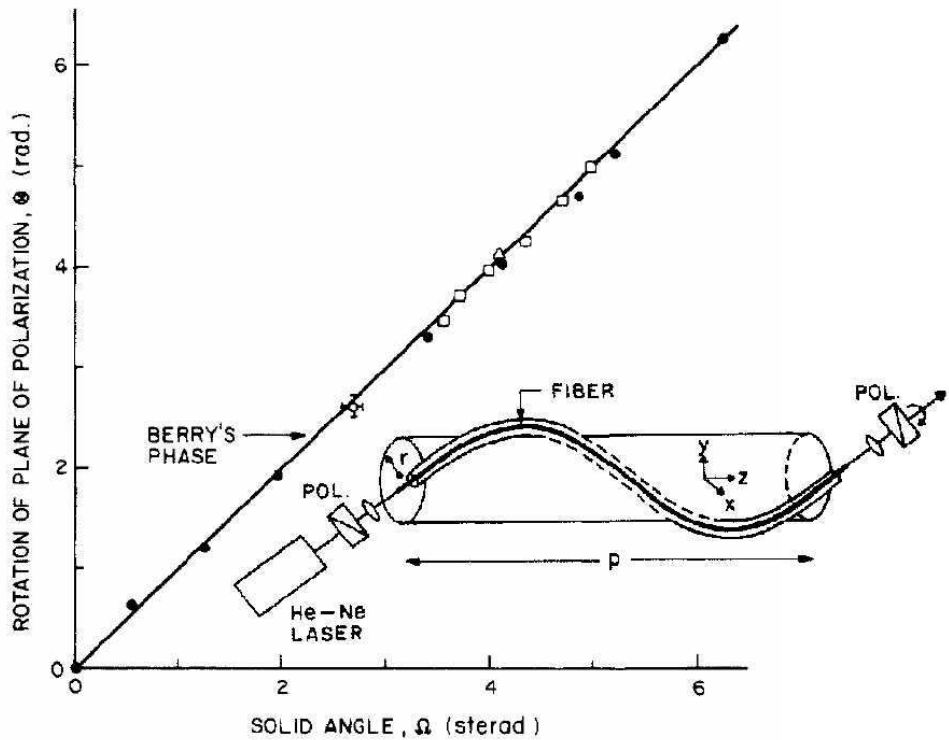,width=0.8\textwidth} \caption{Berry phase
measurement for photons. The inset shows the experimental setup.
Measured angle of rotation of linearly polarized light vs helix
solid angle. Open circles represent the data for uniform helices;
squares and triangle and solid circles represent nonuniform
helices. The solid line is the theoretical prediction based on
Berry's phase. From Ref.~\cite{Tomita1986}.}
\end{center}
\end{figure}

The experiment carried out by Tomita and Chiao \cite{Tomita1986}
is shown in Fig.~6, and shows a very good agreement between
theoretical prediction and experimental results. Note that this
kind of experiment can also be explained entirely from classical
electrodynamics considerations \cite{Berry1987}.

\subsection{Berry phase effects in nuclear magnetic resonance}

\begin{figure}[t]
\label{fig:Suter}
\begin{center}
\epsfig{file=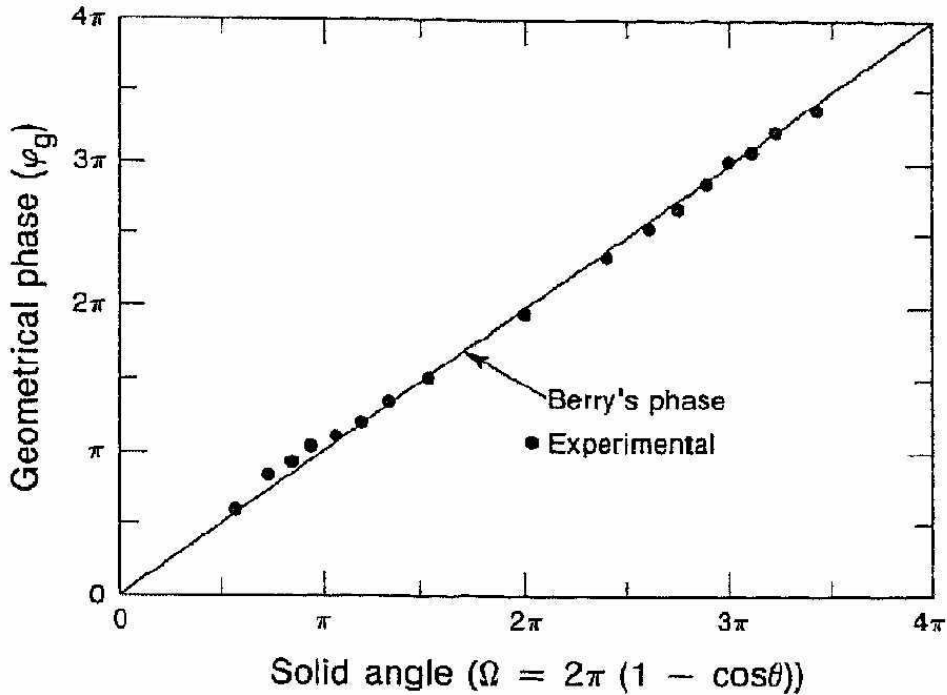,width=0.8\textwidth} \caption{NMR
measurement of the Berry phase. From Ref.~\cite{Suter1987}.}
\end{center}
\end{figure}

Nuclear spins interact very weakly with each other and with their
environment and therefore offer constitute systems that ideally
suited to test the Berry phase of a single spin. The experiment
described below has been performed on protons ($S=1/2$) by Suter
\emph{et al.} \cite{Suter1987} following a proposal of Moody
\emph{et al.} \cite{Moody1986}.

As in a typical nuclear magnetic resonance (NMR) experiment, the
spins are subject to a large, static, orienting field parallel to
the $z$ axis, and to a weak, transverse, field rotating around
$\mathbf{B}_0$ at angular frequency $\omega_\mathrm{RF}$. For
convenience, we express here energies and magnetic fields in units
of angular frequencies, i.e., the Hamiltonian, expressed in the
laboratory frame, reads
\begin{equation}
H(t)= -\omega_L S_z  - \omega_1 [S_x \cos(\omega_\mathrm{RF}t ) +
S_y \sin(\omega_\mathrm{RF}t )] .
\end{equation}
The measured signal is the transverse magnetization $\langle S_x
(t)+ \mathrm{i} \rangle,\langle S_y (t)\rangle $. In the present
case, it is of convenient to perform a transformation from the
laboratory frame to a detector frame, rotating at angular
frequency $\omega_D$. In practice, this is achieved by beating the
measured signal against a reference signal of angular frequency
$\omega_D$. In the detector frame, the Hamiltonian now reads
\begin{equation}
H^\prime(t)= -(\omega_L - \omega_D ) S_z  - \omega_1 [S_x
\cos((\omega_\mathrm{RF} - \omega_D)t) + S_y
\sin((\omega_\mathrm{RF} - \omega_D)t )] .
\end{equation}
Let us define
\begin{equation}
\omega_\mathrm{eff} \equiv \sqrt{(\omega_L - \omega_D)^2 +
\omega_1^2} ,
\end{equation}
which is the magnitude of the total field in the detector frame,
at angle $\theta \equiv \arcsin (\omega_1 / \omega_\mathrm{eff})$
from the $z$ axis and precessing around the $z$ axis at angular
frequency $\omega_\mathrm{RF}- \omega_D$. In the adiabatic limit,
i.e., if $|\omega_\mathrm{RF} - \omega_D | \ll
\omega_\mathrm{eff})$, the adiabatic eigenstates have an energy
$\omega_n = -n\omega_\mathrm{eff}$ ($n = \pm 1/2$). For each cycle
of the effective field around the $z$ axis, the state $n$ acquires
a Berry phase $\gamma_n = -n2\pi (1-\cos \theta )$. Thus if we
prepare the system in a superposition of the two states $n=\pm
1/2$, the Fourier spectrum of the transverse magnetization will
have a component of angular frequency $\omega_\mathrm{eff} +
(\omega_\mathrm{RF}- \omega_D)2\pi (1-\cos \theta )$, where the
last term arising from the Berry phase, as shown in Fig.~7.

\section{Berry phase for itinerant electrons in a solid}

\subsection{General formulation}

We now want to discuss the Berry phase of electrons in solids. Let
us consider (non-interacting) electrons subject to a scalar
potential and to a Zeeman (or exchange) field, whose direction is
spatially nonuniform. As they move through the solid, the
electrons experience, in their proper reference frame, a Zeeman
field whose direction changes with time. If this change is slow
enough, the electron spin has to follow it adiabatically and
therefore accumulates a Berry phase \cite{Loss1990, Stern1992,
Loss1992}.

Before discussing the resulting physical consequences, let us
formulate the problem more precisely. I follow here the discussion
of Ref.~\cite{Bruno2004}. The corresponding Hamiltonian is
\begin{equation}
H=-\frac{\hbar^2}{2m}\frac{\partial^2}{\partial\mathbf{r}^2} +
V(\mathbf{r}) - \Delta(\mathbf{r}) \mathbf{n(r)}\cdot \bm{\sigma}
.
\end{equation}
We use a gauge transformation $T({\bf r})$, which makes the
quantization axis oriented along vector ${\bf n}({\bf r})$ at each
point. It transforms the last term in the above equation as
$T^\dag ({\bf r})\, \left[ \bm{\sigma} \cdot {\bf n}({\bf
r})\right] \,T({\bf r})=\sigma _z$, corresponding to a local
rotation of the quantization axis from $z$ axis to the axis along
${\bf n}({\bf r})$. The transformed Hamiltonian describes the
electrons moving in a (spinor) gauge potential ${\bf A}({\bf r})$,
\begin{equation}
H^\prime \equiv T^\dag H T =-\frac{\hbar ^2}{2m}\, \left(
\frac{\partial }{\partial {\bf r}} -\frac{ie}{\hbar c}\; {\bf
A}({\bf r}) \right)^2 + V(\mathbf{r}) -\Delta (\mathbf{r}) \sigma
_z\, ,
\end{equation}
where $A_i({\bf r})=-2\pi i\, \phi _0\, T^{\dag }({\bf r})\,
\partial _i\, T({\bf r})$, $\phi _0=hc/\left| e\right| $
is the flux quantum. For convenience, we have defined the gauge
potential ${\bf A}({\bf r})$ to have the same dimension as the
electromagnetic vector potential. The components of ${\bf A}({\bf
r})$ can be found easily using an explicit form of $T({\bf r})$.

The above Hamiltonian with the matrix ${\bf A}({\bf r})$ contains
terms coupling the two spin states. If the rate at which the
spin-quantization axis varies is slow enough as compared to the
Larmor precession frequency (as seen from the moving electron's
frame), the spin will follow adiabatically the local
spin-quantization axis, and these spin-flip terms can be
neglected. The variation rate of the spin quantization axis is
(for an electron at the Fermi level) $v_F/\xi$, where $\xi$ is the
characteristic length of variation of the spin-quantization axis,
so that the adiabaticity condition is
\begin{equation}
\frac{\hbar v_F}{\xi} \ll \Delta .
\end{equation}
With this approximation we now obtain
\begin{equation}
H^\prime \simeq -\frac{\hbar ^2}{2m}\; \left( \frac{\partial
}{\partial {\bf r}}-\frac{ie}{\hbar c} \mathbf{a} ({\bf r})
\sigma_z \right)^2 + V({\bf r}) + V^\prime(\mathbf{r}) -\Delta
(\mathbf{r}) \sigma _z ,
\end{equation}
where
\begin{equation}
\label{4} a_i({\bf r})\equiv \frac{\pi \phi _0\, \left( n_x\,
\partial _in_y-n_y\, \partial _in_x\right)} {1+n_z}
\end{equation}
is an effective vector potential arising from the Berry phase, and
\begin{equation}
V^\prime({\bf r})=(\hbar ^2/8m)\, \sum_{i,\mu} (\partial
_in_\mu)^2
\end{equation}
is an effective scalar potential. Since the Hamiltonian is
diagonal in spin, we can treat the two spin subbands separately.
For each of the spin subband, we have mapped the original problem
on the simpler one of a spinless electron moving in effective
scalar and vector potentials. The effective vector potential in
turn gives rise to an effective magnetic field $\mathbf{b}\equiv
\nabla \times \mathbf{a}$, whose components are expressed in terms
of the magnetization texture as
\begin{equation}
B_i =\frac{\phi_0}{4\pi }\; \varepsilon_{ijk}
\varepsilon_{\mu\nu\lambda}\, n_\mu \, (\partial_j n_\nu )\,
(\partial_k n_\lambda ) .
\end{equation}

The physical consequences of the effective vector potential and
effective magnetic field are exactly the same as those of the
familiar electro-magnetic counterparts, and can be classified in
two different classes:
\begin{enumerate}
\item local effects such as the Lorentz force. These effects are
classical in origin (see the illuminating discussion on this point
given by Aharonov and Stern \cite{Aharonov1992}), and therefore do
not require phase coherence.
\item non-local interference effects, such as Aharonov-Bohm-like
effects and persistent currents. These effects are intrinsically quantum
mechanical, and require phase coherence.
\end{enumerate}

\subsection{Anomalous Hall effect due to the Berry phase in a textured ferromagnet}

The Hall effect consists in the appearance of a voltage transverse
to the current in a conducting system. As it is antisymmetric with
respect to time reversal, it may appear only in the presence of a
term in the Hamiltonian breaking time reversal invariance. Until
recently, two different mechanism were recognize to give rise to
the Hall effect:
\begin{enumerate}
\item the electro-magnetic Lorentz force due to a usual magnetic
field; in the classical regime (normal Hall effect), this is well
described by the Drude theory; in the quantum limit, the
spectacular quantum Hall effect is obtained.
\item in absence of an external magnetic field, time-reversal
invariance is also broken if the system exhibits magnetic order.
However, this fact is not enough to induce a Hall effect if the
magnetic order is uniform and spin-orbit coupling is absent or
negligible, because the system is then invariant by under a global
rotation of $\pi$ of the spins, which is equivalent to time
reversal in this case (uniform magnetization). Therefore Hall
effect arises only as a consequence of simultaneous presence of
spontaneous magnetic order and spin-orbit coupling. This mechanism
is called the anomalous Hall effect.
\end{enumerate}

Recently, however, it was realized that a third mechanism could
give rise to the Hall effect in ferromagnetic, in absence of an
external magnetic field, and of the spin-orbit coupling, if the
magnetization is non-uniform and exhibits a non-trivial texture
\cite{Matl1998, Ye1999, Ohgushi2000, Chun2000, Taguchi2001,
Taguchi2001b, Lyanda-Geller2001, Nagaosa2001, Yanagihara2002,
Tatara2002, Onoda2003, Taguchi2003, Bruno2004}.

The central idea is the following: as we have discussed above that
if the spin-orbit coupling is negligible, the system is invariant
under a global rotation of the magnetization. However, if the
magnetization is non-uniform, a rotation of $\pi$ is generally not
equivalent to a time-reversal, so that Hall effect may arise. At
the microscopic level, the origin of the Hall effect in such a
case is the effective Lorentz force due to the Berry phase of the
magnetization texture.

\begin{figure}[t]
\label{fig:Taguchi1}
\begin{center}
\epsfig{file=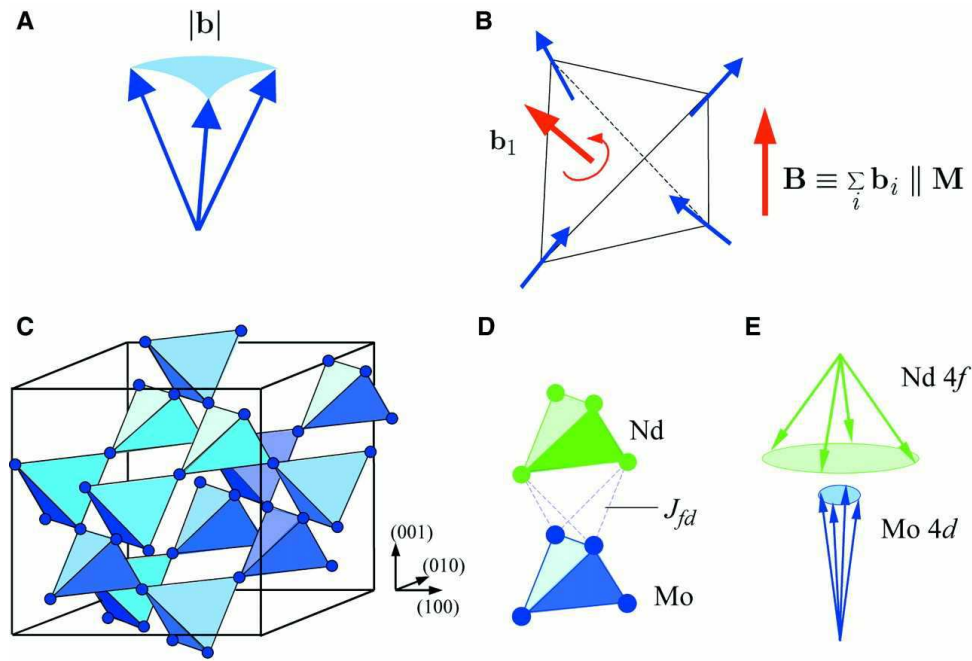,width=0.8\textwidth} \caption{Schematic
magnetic and crystal structures of pyrochlore. (A) Spin chirality,
that is, the solid angle subtended by the three spins.
(B)``Two-in, two-out'' spin structure, in which each spin points
along the line that connects the center of the tetrahedron and the
vertex. The total fictitious magnetic field is the vector sum of
each fictitious magnetic flux that penetrates each plaquette. (C)
The B sublattice of pyrochlore structure A$_2$B$_2$O$_7$. The A
sublattice is structurally identical with this one, but is
displaced by half a lattice constant. (D) Relative position of Nd
tetrahedron (green circles) and Mo tetrahedron (blue circles) in
Nd$_2$Mo$_2$O$_7$ pyrochlore. (E) The "umbrella" structure
observed for Nd$_2$Mo$_2$O$_7$ (A$=$Nd, B$=$Mo) by a neutron
diffraction study. From Ref.~\cite{Taguchi2001}.}
\end{center}
\end{figure}

We shall discuss below as an example the results of Taguchi
\emph{et al.} \cite{Taguchi2001}. In this work the authors
investigated the compound Nd$_2$Mo$_2$O$_7$, which has the
pyrochlore structure shown on Fig.~8. Due to their large
spin-orbit coupling the Nd 4f moments of a given tetrahedron adopt
the ``two-in, two-out'' structure shown on Fig.~8B. The Mo 4d
moments which are ferromagnetically coupled to each other and
antiferromagnetically to the the Nd moments therefore adopt and
non-collinear umbrella structure, whose chirality gives rise to
the anomalous Hall effect: as electrons moves on a triangle face
of a tetrahedron, they acquire a Berry phase, and experience the
associated Lorentz force. This mechanism is dominant at low
temperature, where other mechanisms due to the spin-orbit coupling
(giving contributions to the Hall resistivity linear or quadratic
in the longitudinal resistivity) are strongly suppressed. That the
origin of the anomalous Hall effect is the Berry phase due to the
texture is further indicated by noting that the application of a
magnetic field reduces the solid angle of the umbrella texture and
hence the Berry phase and the associated Hall effect.

\begin{figure}[t]
\label{fig:Taguchi3b}
\begin{center}
\epsfig{file=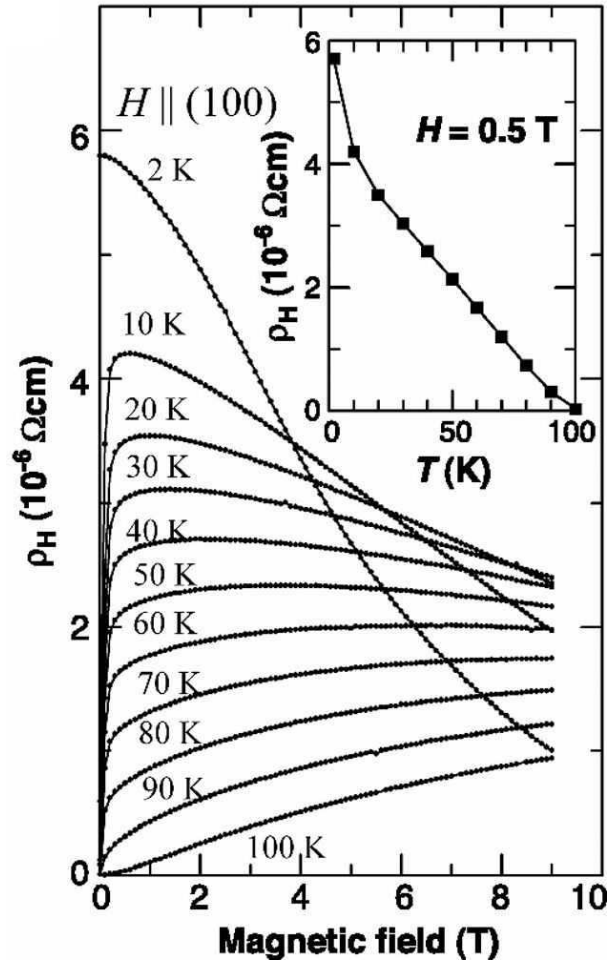,width=0.5\textwidth} \caption{Magnetic field
dependence of the Hall resistivity $\rho_H$ with $H \| (100)$ at
several temperatures. The inset shows the temperature dependence
of $\rho_H$ at 0.5~T, which is a measure of the anomalous Hall
term. The $\rho_H$ at a low field ($< 0.3$~T) is finite at 2~K,
whereas it tends to zero above 10~K, in accord with the presence
or absence of remnant magnetization at the respective
temperatures. From Ref.~\cite{Taguchi2001}.}
\end{center}
\end{figure}

\begin{figure}
\label{fig:Ohgushi}
\begin{center}
\epsfig{file=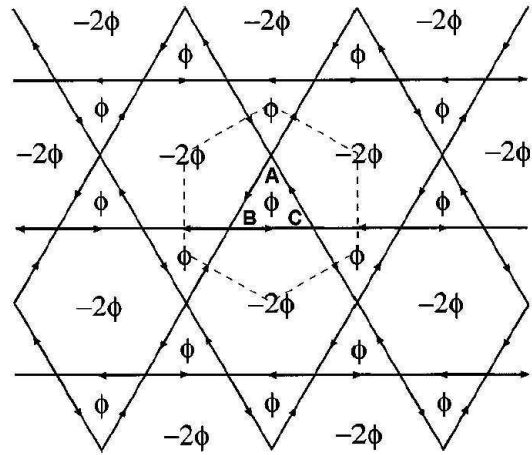,width=0.5\textwidth} \caption{Kagom\'{e}
lattice. The dotted line represents the Wigner-Seitz unit cell,
which contains three independent sites $A,B,C$. From
Ref.~\cite{Ohgushi2000}.}
\end{center}
\end{figure}

One should note, however, that the average effective magnetic
field due to the Berry phase is zero on the present case. This can
be understood by noting \cite{Ohgushi2000} that the Mo planes
perpendicular to (111) axes are kagom\'{e} lattices, with a Berry
phase of $+\varphi$ on the triangles and $-2\varphi$ on the
hexagons. Since there are two triangles and one hexagon per unit
cell the effective magnetic field due to the Berry phase is zero
on average. However, since the circuits with positive (triangles)
and negative (hexagons) Berry phase are inequivalent, time
reversal invariance is still broken, and a non zero Hall effect
may (and generally does) result \cite{Ohgushi2000}. However,
because the effective field due to the Berry phase vanishes on
average, the resulting Hall effect is considerably weaker than
that one would obtain for a non-zero net effective field.

\begin{figure}[p]
\label{fig:Bruno1}
\begin{center}
\epsfig{file=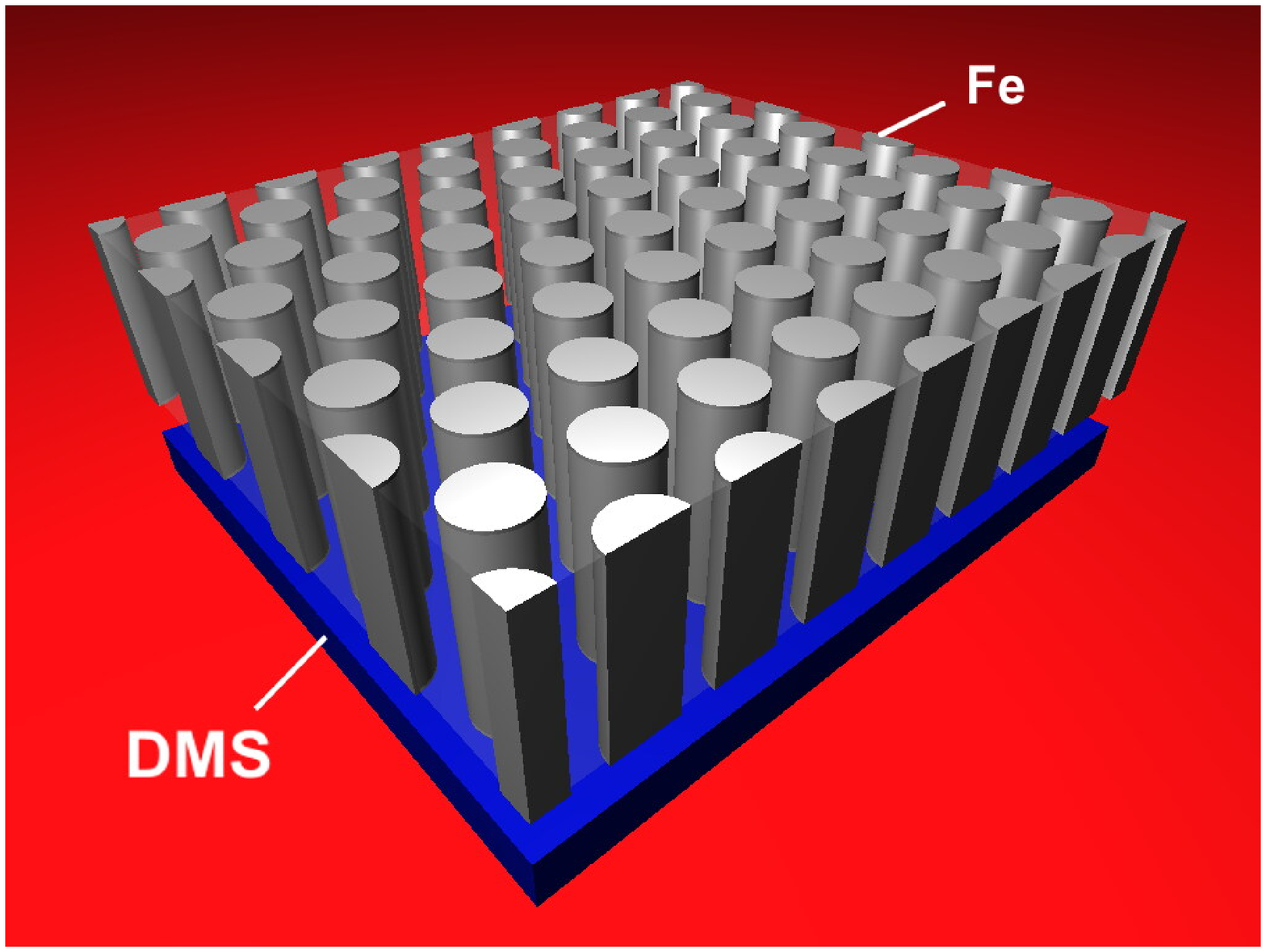,width=0.45\textwidth} $\ $
\epsfig{file=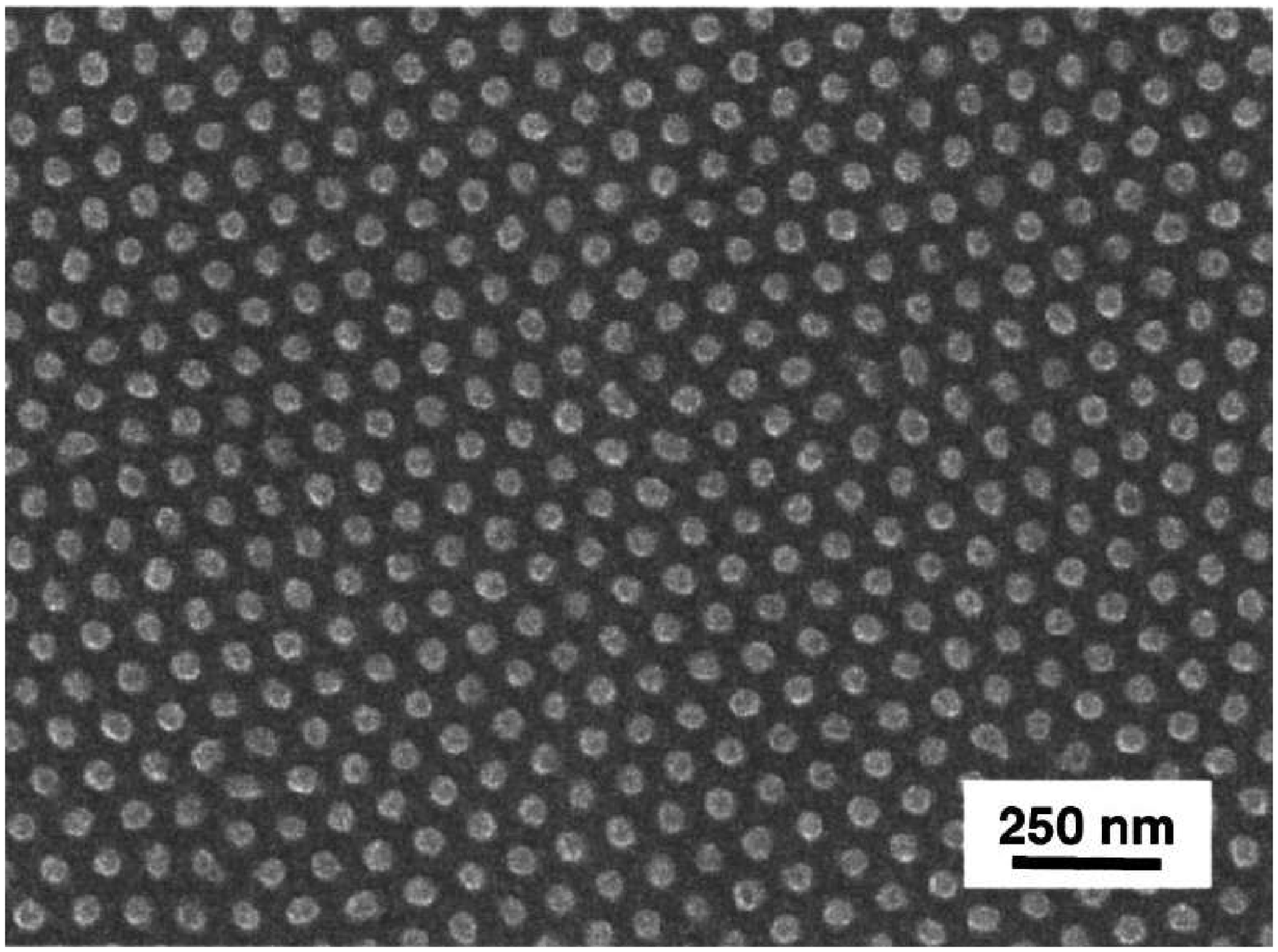,width=0.45\textwidth}\caption{Left:
The proposed structure consisting of a triangular lattice of
magnetic nanocylinders on top of 2D diluted magnetic
semiconductor. From Ref.~\cite{Bruno2004}. Right: Example of an
triangular array of Ni nanocylinders (cylinder distance $=$100~nm)
in an alumina matrix. From Ref.~\cite{Nielsch2001}.}
\end{center}
\end{figure}

\begin{figure}[p]
\label{fig:Bruno2}
\begin{center}
\epsfig{file=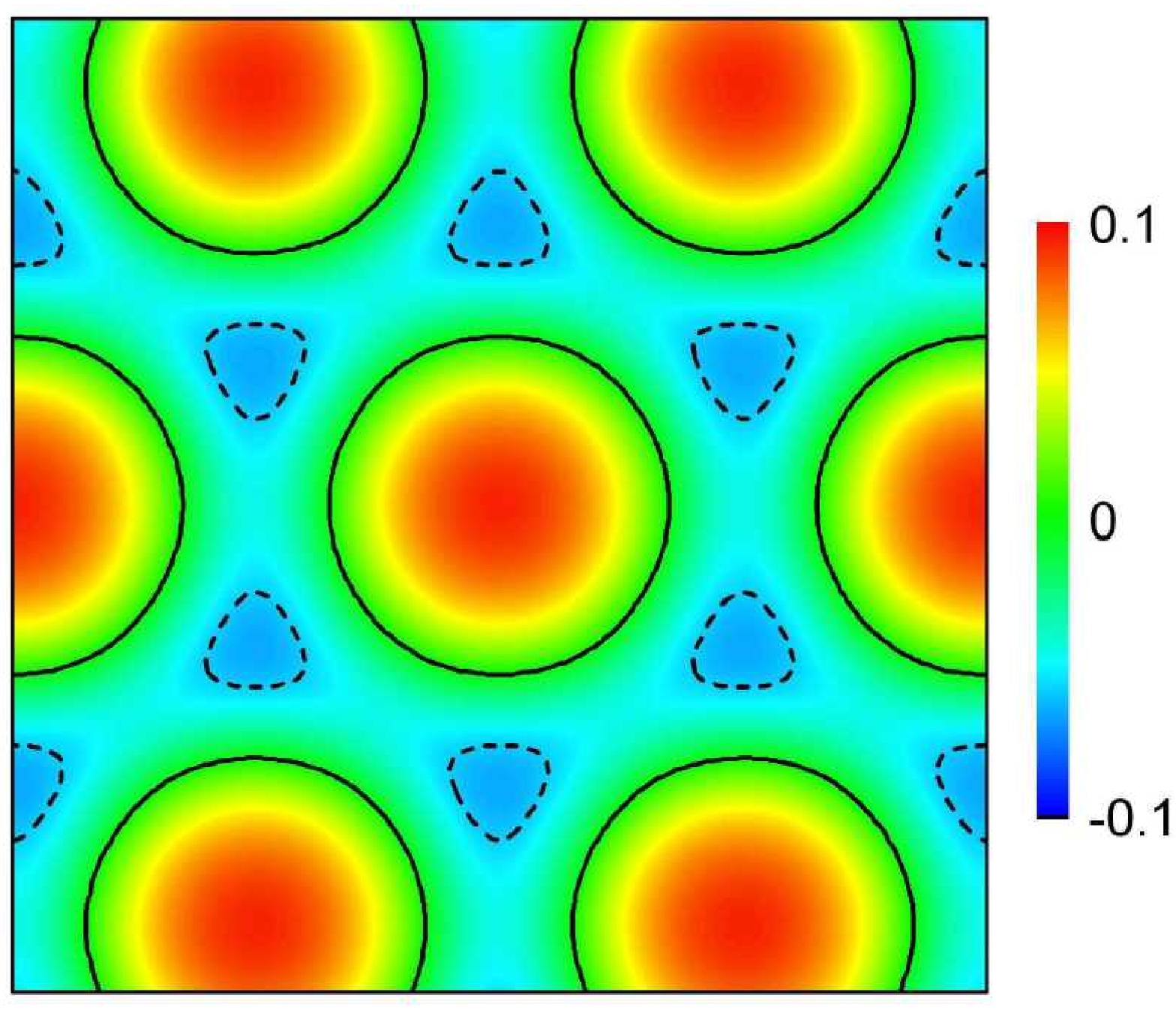,width=0.48\textwidth}
\epsfig{file=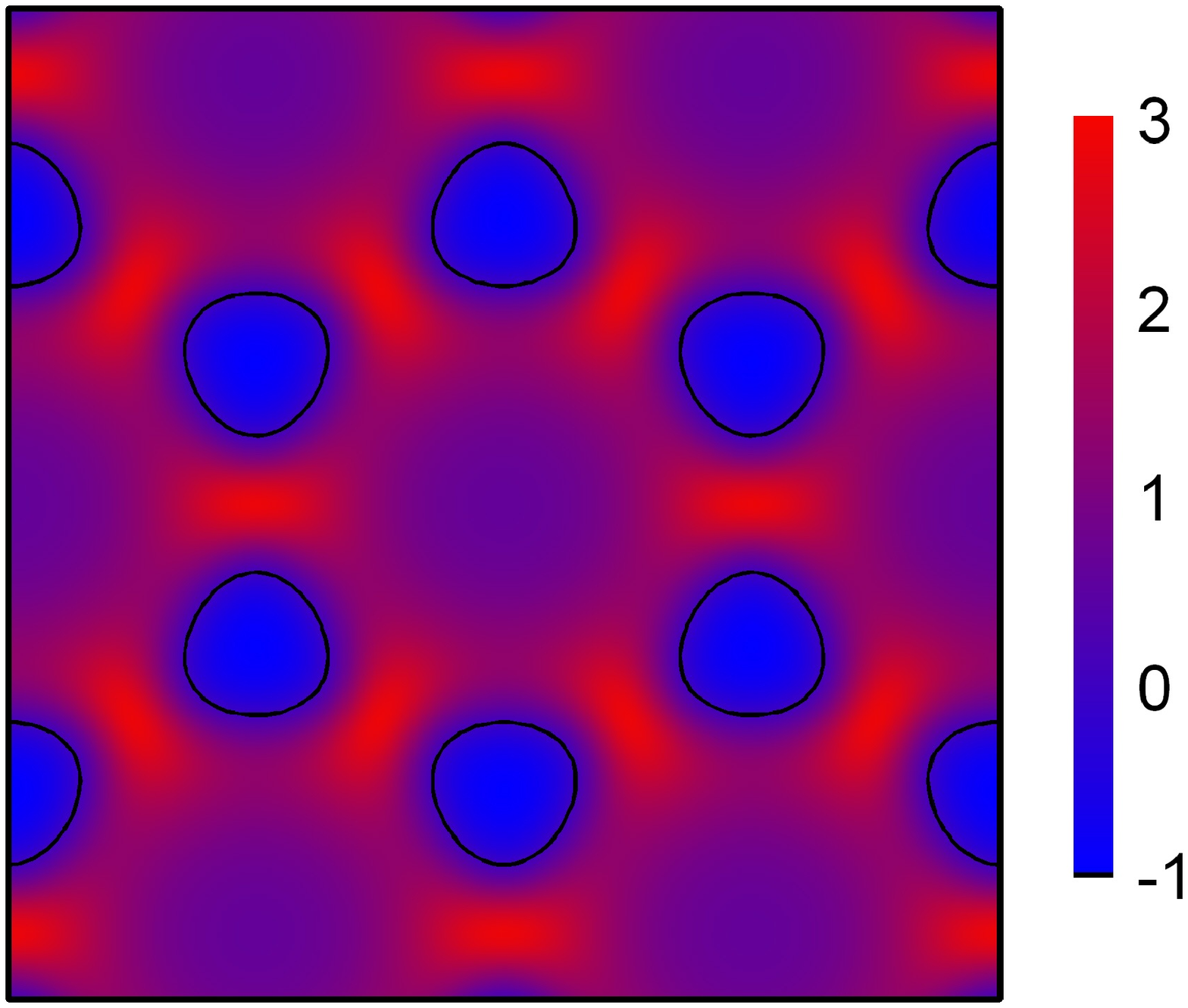,width=0.48\textwidth}\caption{Left:
Distribution of the $z$-component of dipolar field $B/4 \pi M_s$
inside the semiconductor film for the triangular lattice of
magnetic nanocylinders, for a zero external field. The black solid
circles correspond to the lines with $B_z = 0$. Dashed lines
correspond to the lines with $B_z = 0$ under a uniform external
magnetic field $B_\mathrm{ext}/4\pi M_s =0.058$. Right:
Topological field $B_t(\mathbf{r})$ (in units of $\phi_0$ per unit
cell area) for the triangular lattice of magnetic nanocylinders.
From Ref.~\cite{Bruno2004}.}
\end{center}
\end{figure}

One should also notice that, as in the example discussed above,
spin-orbit coupling always has to be invoked in order to obtain a
spin-texture with a definite chirality and, hence, a non-zero Hall
effect \cite{Ye1999, Chun2000, Lyanda-Geller2001, Tatara2002,
Onoda2003}. This is, however, quite different from the mechanisms
of Hall effect in which the spin-orbit coupling directly
influences the motion of the conduction electrons.

Until recently, all cases of anomalous Hall due to a chiral spin
texture discussed in the literature considered a texture at the
atomic level. Recently, it has been proposed that a mesoscopic
scale magnetic field texture can be produced by using magnetic
nanostructures \cite{Bruno2004}. For example, one can consider an
array of magnetic nanocylinders (all magnetized in the same
direction along their axis), as shown in Fig.~11 (left panel), in
order to generate a non-uniform dipolar stray field in a
two-dimensional electron (or hole) gas placed just underneath.
Such structure can be fabricated by electrolytic techniques, as
shown on Fig.~11 (right panel).

\begin{figure}[t]
\label{fig:Topo_flux}
\begin{center}
\epsfig{file=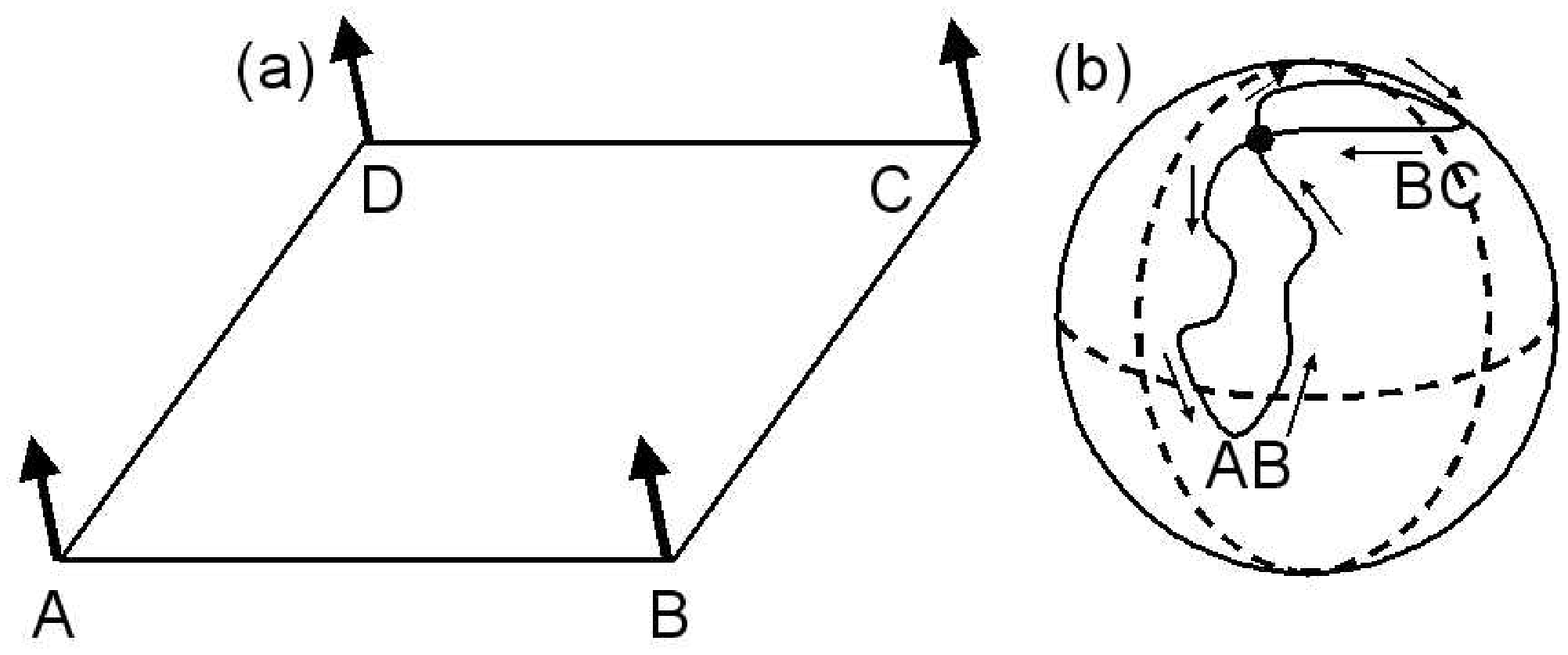,width=0.7\textwidth}\caption{(a) unit cell
in real space; (b) paths AB and BC on the sphere of unit radius.}
\end{center}
\end{figure}

The Fig.~12 shows the $z$-component of the dipolar stray field
(left panel) and topological field generated by the Berry phase
(right). It is noteworthy that for a lattice of Fe cylinders with
a pitch of 100~nm, the dipolar field at a distance 20~nm
underneath (the average value of which is always zero) has a
maximum absolute value of about 2~kG, whereas the topological
field has an non-zero average value of about 5~kG, with local
values ranging between $-5$ and $+15$~kG. Thus, one sees that a
rather weak dipolar stray field averaging to zero generates in the
two-dimensional electron gas a much stronger topological field
with a non-zero average!

The topological field has a number of interesting properties:
\begin{itemize}
\item
First, one can show that the total flux of the topological field
through a unit cell is always an integer multiple of the flux
quantum. To see this, let us consider the Berry phase
corresponding to circuit going around a unit cell (e.g., along the
path ABCDA on Fig.~13 (left panel). Because of the translational
periodicity of the system, the local field points along the same
direction at the four corners A, B, C, D of the unit cell, and
thus correspond to the same point (represented by the solid dot)
on the sphere of unit radius on which the corresponding path for
the field direction takes place; the path ABC is shown, and one
sees immediately that, because of the translational periodicity,
the path CDA is exactly the same, described in the reversed
direction, so that the Berry phase corresponding to the path ABCDA
is zero modulo $2\pi$ which implies the quantization of the total
flux in units of $\phi_0$. This constitutes an example of
topological quantization.
\item
Next, let us try to understand what is the actual integer value
taken by the total topological flux. The above reasoning does tell
us anything about it, because we don't know how many times we are
wrapping the sphere when integrating over the unit cell. To know
this, we may consider the closed lines where the $z$-component of
the stray field vanishes. This lines corresponds to paths going an
integer number of times around the ``equator'' of the unit sphere.
Such lines are shown on Fig.~12 (left panel, solid lines). One can
then see that each round trip around the equator contributes to
one flux quantum. In counting this, one should be careful in
getting the sign correctly. In the case shown in Fig.~12 for a
vanishing external field, one obtains that the total topological
flux per unit cell is $+\phi_0$.
\item
Finally, one can see that the lines of zero $z$-component of the
stray field will change if one applies an external magnetic field.
For example, in the case considered here, under application of a
uniform external field along the magnetization direction of the
nanocylinders, the regions of positive $z$-field will expand, so
that the topology of the lines of zero $z$-field will eventually
change. This is shown by the dashed lines on the left panel of
Fig.12 ($B_\mathrm{ext}/4\pi M_s =0.058$). From the above
discussion, this implies a change in the topological flux per unit
cell, which now takes the value $-2\phi_0$ (the factor 2 is
because there are 2 lines per unit cell, and the factor $-1$ is
because the circulation is reversed.
\end{itemize}

\begin{figure}[t]
\label{fig:Bruno4}
\begin{center}
\epsfig{file=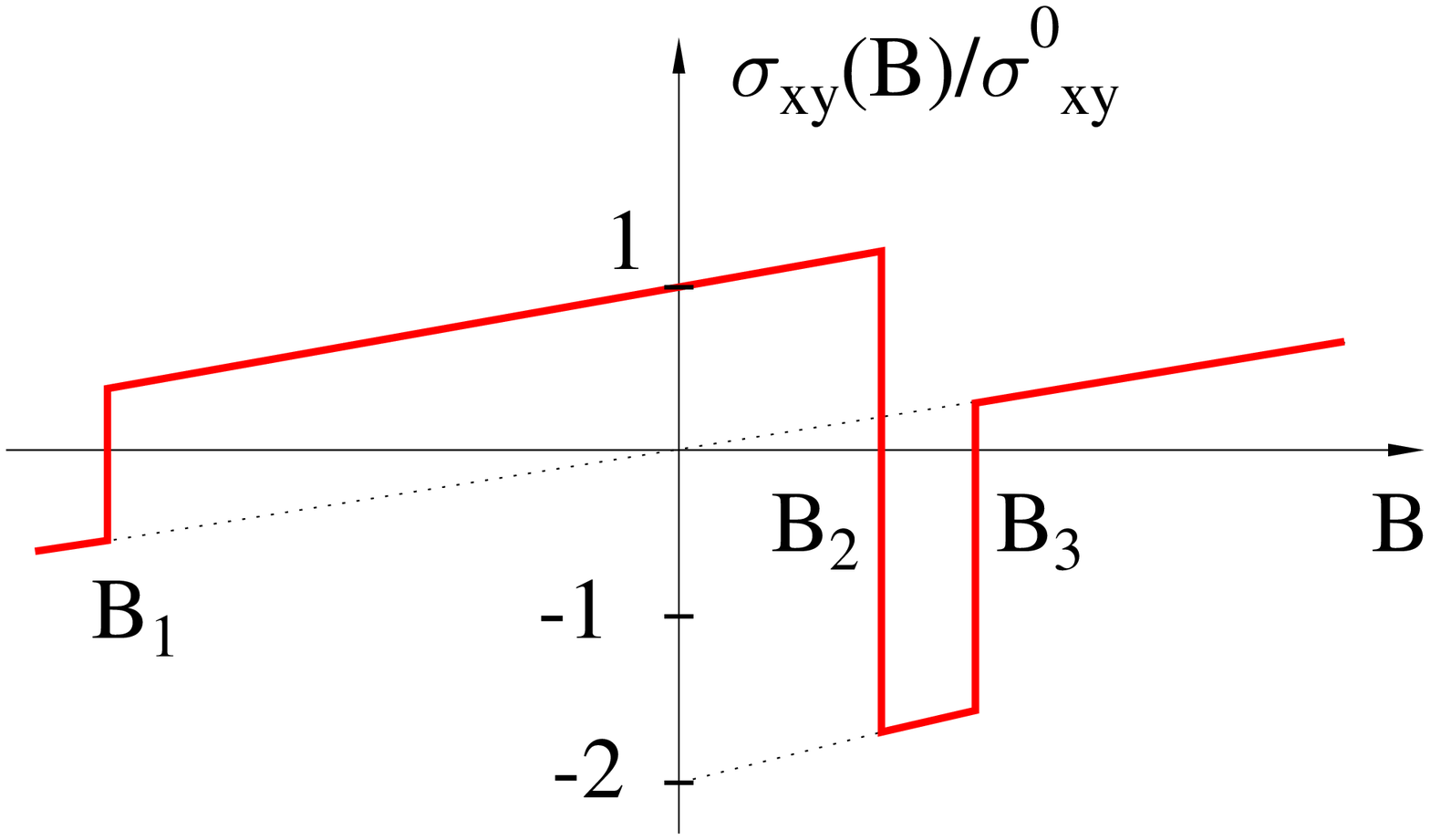,width=0.5\textwidth}\caption{Dependence of
the Hall conductivity on external magnetic field (schematically).
The slope corresponds to the contribution of the normal Hall
effect; $\sigma^0_{xy}$ is the Hall conductivity corresponding to
a topological flux per unit cell equal to $\phi_0$. From
Ref.~\cite{Bruno2004}.}
\end{center}
\end{figure}

The properties discussed above indicate that by application an a
rather small uniform external field, one can change the average
value of the topological field. This will give rise to changes in
the anomalous Hall effect. As a first approximation, one can
estimate the Hall effect by using the Drude model and by
neglecting the spacial fluctuations of the dipolar and topological
fields. In this case, the Hall effect is just given by the
familiar Drude formula, with an effective field equal to the sum
of the external magnetic field and the average topological field.
If the two spin subbands contribute, one simply has to sum the
contributions of the two subbands, taking into account the fact
that the topological field is of opposite sign for the two spin
subbands. For the case discussed here, and assuming that only one
spin subband is occupied, the resulting behavior of the Hall
effect is shown on Fig.~14. For the chosen parameters, he values
of the critical fields are $B_1\simeq -2$~kG, $B_2\simeq 0.9$~kG,
and $B_3\simeq 1.3$~kG. The uniform slope comes from the normal
Hall effect (Lorentz force of the external magnetic field),
whereas the remaining non-monotonous contribution arises from the
Lorentz force due to the topological field of the Berry phase.
Such a characteristic non-monotonous behavior would constitute a
signature of the Berry phase contribution of the Hall effect and
can be tested experimentally. One should point out, however, that
in the vicinity of the critical fields where the topological flux
abruptly changes, the adiabaticity condition cannot be well
satisfied, so that, in practice, a rounded curve would be
obtained.

In order to identify a system for which the above predictions can
be satisfied, we have to look for a system with a large Zeeman
splitting, in order to satisfy as well as possible the condition
of adiabaticity. We propose to use II–VI dilute magnetic
semiconductors (DMS) which exhibit giant Zeeman splitting; p-type
Mn doped DMS are best suited since the exchange constants for
holes are much larger than for electrons \cite{Furdyna1988,
Kossut1993}. For a detailed discussion see Ref.~\cite{Bruno2004}.

Before closing this section on the anomalous Hall effect, I wish
to point that, while the Berry phase allowed to identified a new
mechanism of Hall effect arising from the chirality of the spin
texture, in absence of an external field and of the spin-orbit
coupling, as we have discussed above, it also allows to give a
modern interpretation to the previously known mechanisms for the
Hall effect (normal Hall effect, classical or quantized; anomalous
Hall effect), due either to an external magnetic field, or to the
spin-orbit coupling. In this case, one deals with Berry phase in
momentum space instead of real space as discussed above. For a
detailed discussion, see Refs.~\cite{Simon1983, Kohmoto1985,
Niu1985, Kohmoto1993, Chang1996, Sundaram1999, Jungwirth2002,
Onoda2002, Onoda2003b, Fang2003, Culcer2003, Yao2004,
Haldane2004}.

\subsection{Interference effects due to the Berry phase in an
Aharonov-Bohm ring}

\begin{figure}[t]
\label{fig:Loss}
\begin{center}
\epsfig{file=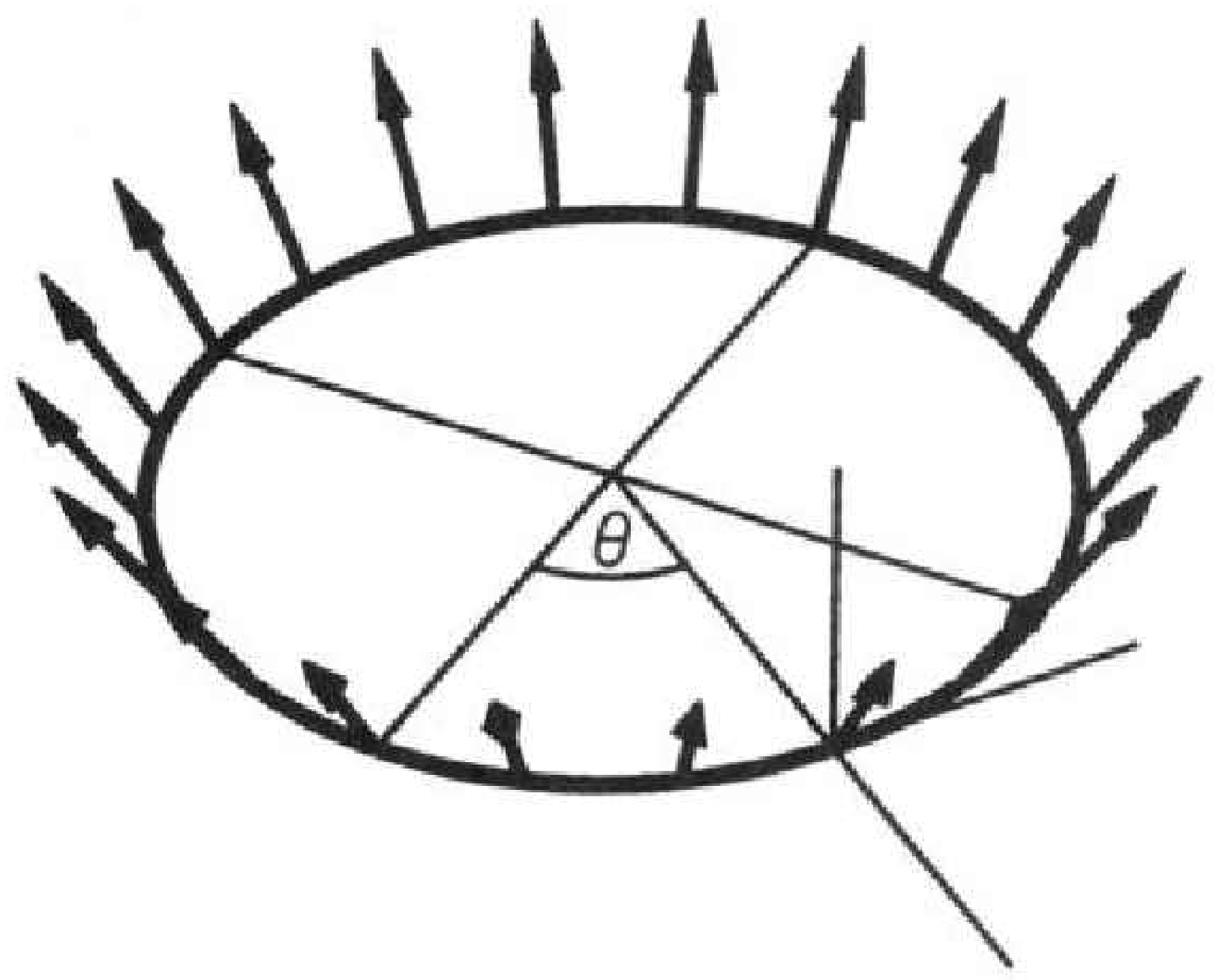,width=0.5\textwidth}\caption{Texture of the
magnetic field (or magnetization) in a metallic ring. From
Ref.~\cite{Loss1992}.}
\end{center}
\end{figure}

As mentioned earlier, the Berry phase accumulated by electrons
moving in a non-trivial magnetic texture can give rise to
interference effects, of which the archetype is the Aharonov-Bohm
effect. It has been proposed by Loss \emph{et al.} \cite{Loss1990,
Loss1992} and by Stern \cite{Stern1992} that a metallic ring
subject to a textured magnetic field (or  magnetization) as
depicted in Fig.~15 would yield a Berry phase for an electron
moving around the ring, and hence a dependence of the conductance
of the ring (when connected to current leads) upon the solid angle
described by the magnetization \cite{Stern1992, Stern1997}, as
well as to persistent charge and spin currents (for a
non-connected ring) \cite{Loss1990, Loss1992}.

\begin{figure}[t]
\label{fig:Rashba}
\begin{center}
\epsfig{file=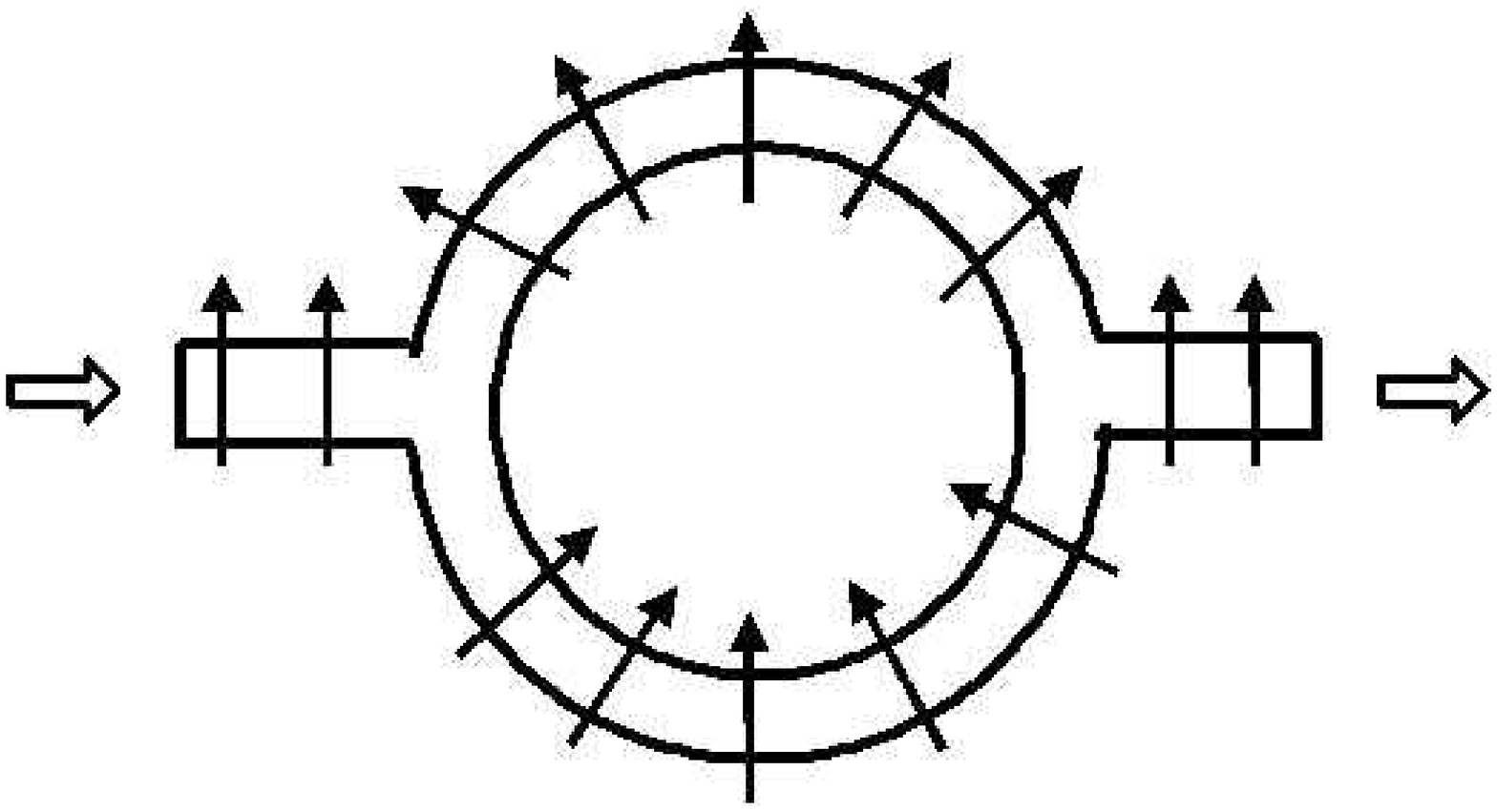,width=0.5\textwidth}\caption{Sketch of the
effective field due to the Rashba effect for an electron moving
from the left lead to the right lead.}
\end{center}
\end{figure}

So far, it has not been possible to test experimentally this
prediction in the configuration described above (i.e., by using a
textured magnetic field or magnetization). However, several
authors \cite{Aronov1993, Qian1994} have indicated that a similar
Berry phase may be obtained by using the combined effect of the
Zeeman coupling to a uniform magnetic field (parallel to the ring
axis) and of Rashba-type spin-orbit coupling \cite{Rashba1960,
Bychkov1984}, as described by the following Hamiltonian
\begin{equation}
H = \frac{\mathbf{p}^2}{2m} - B\sigma_z + \alpha (\mathbf{p}\times
\bm{\sigma} )\cdot \mathbf{\hat{z}} + V(\mathbf{r}).
\end{equation}
In the above equation, the third term gives the Rashba spin-orbit
coupling, while the last one confines the electron to the ring.
The Rashba effect acts as an effective magnetic field
perpendicular to the plane and to the direction of motion.

\begin{figure}[t]
\label{fig:Mopurgo}
\begin{center}
\epsfig{file=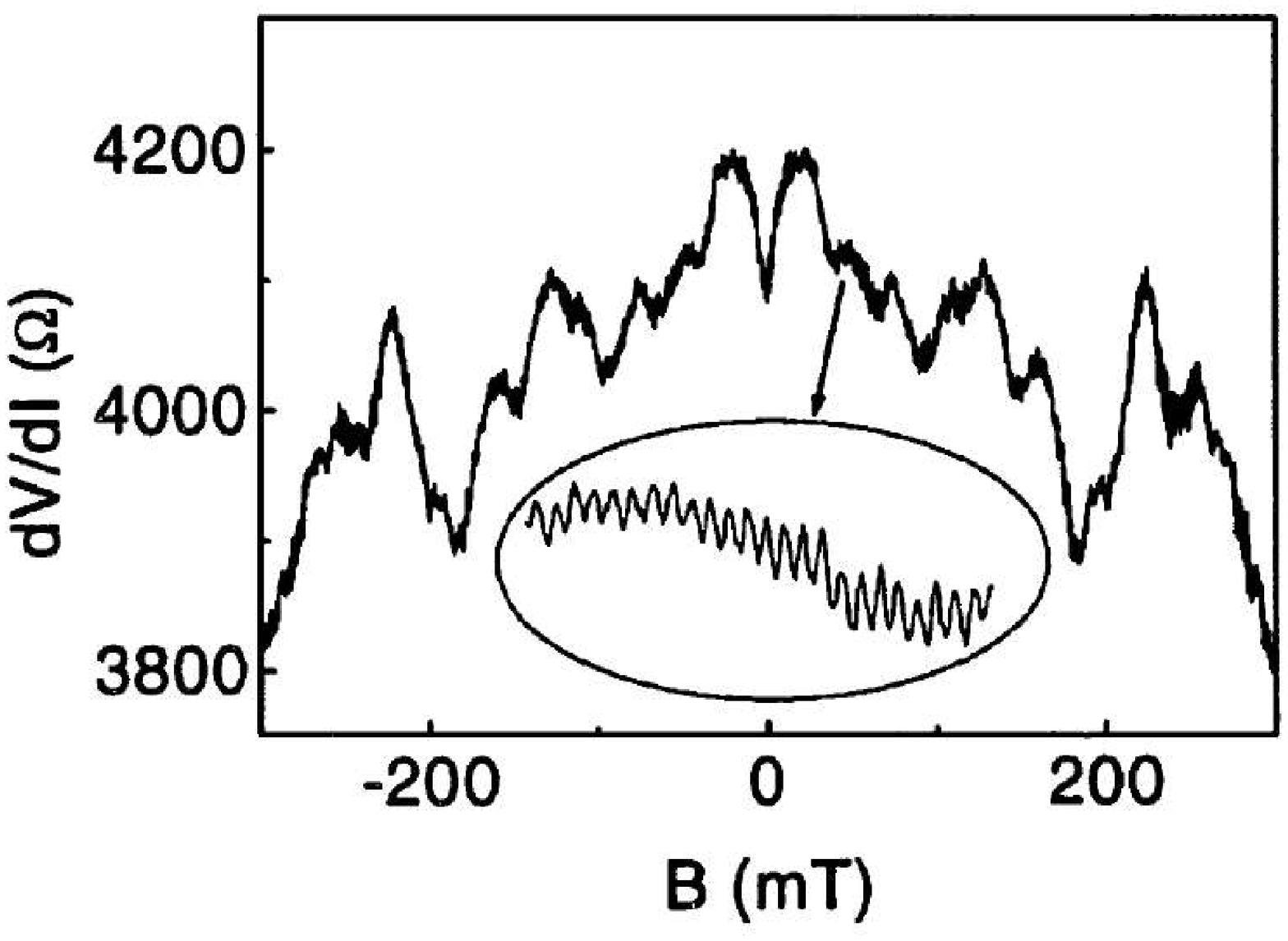,width=0.48\textwidth}
\epsfig{file=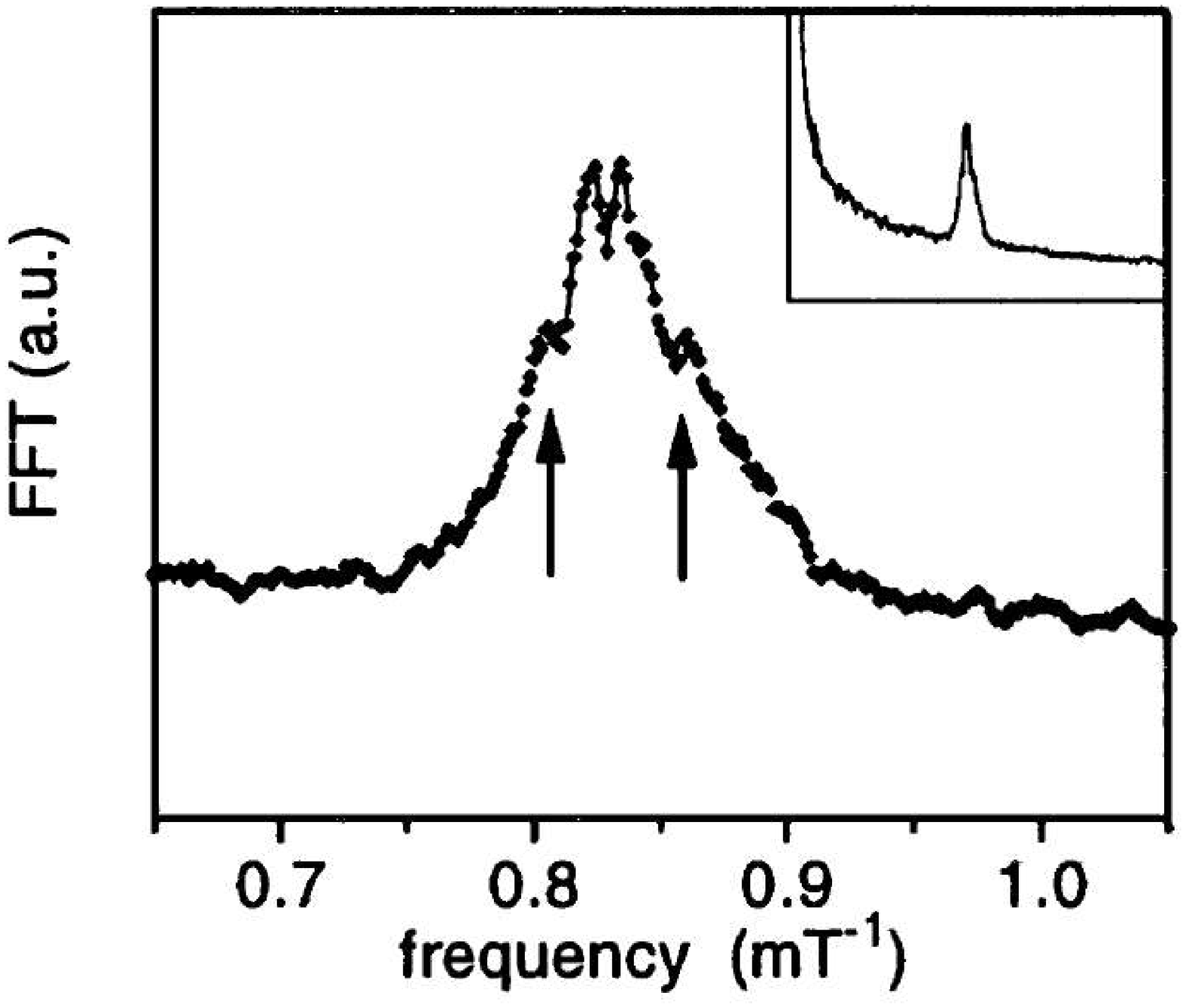,width=0.48\textwidth}\caption{Left
panel: average of about 30 curves $R(B)$ (the inset is an
enlargement of the small part of the curve). Right panel: the peak
of the average Fourier spectrum: the splitting is evident, as well
as some structure on the sides (pointed by the arrows). The inset
shows the same curve on a larger frequency range. From
Ref.~\cite{Mopurgo1998}}
\end{center}
\end{figure}

There is, however, a crucial difference with respect to a true
magnetic field, namely that the Rashba term is invariant under
time reversal (unlike a true magnetic field), which is manifest
from the fact the associated effective field changes sign as the
motion is reversed. Because of this, there is no phase shift from
the Berry phase between the paths going through the upper and
lower arms of the ring, as sketched on Fig.~16, unlike what would
be expected for a real magnetic field as in Fig.~15. Therefore,
Aharonov-Bohm like interferences in this configuration have to
involve paths that wind completely the ring a different number of
times. The associated Berry phase will of course be superimposed
to the usual Aharonov-Bohm phase and therefore modify the
Aharonov-Bohm oscillations of the ring conductance versus magnetic
field. Such observations, indicating the presence of the Berry
phase, have been made by various groups \cite{Mopurgo1998,
Yau2002, Yang2004}. The results of Mopurgo \emph{et al.}
\cite{Mopurgo1998} are shown in Fig.~17, where the signature of
the Berry phase is given by the splitting of the $e/h$ peak in the
average Fourier spectrum.

\section{Further effects of Berry phase in magnetism}

\begin{figure}[t]
\label{fig:Wernsdorfer}
\begin{center}
\epsfig{file=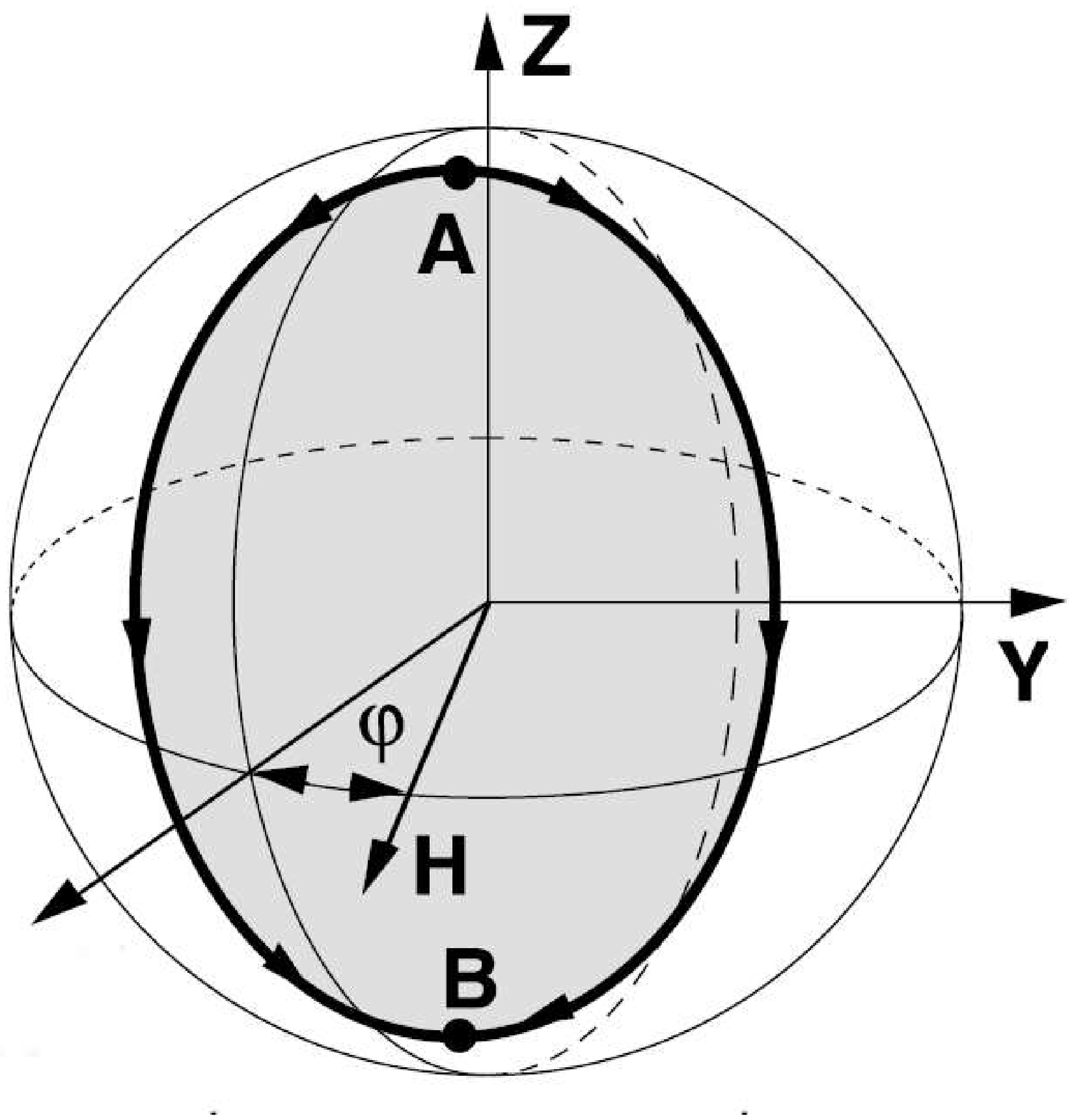,width=0.5\textwidth} \caption{Sketch of the
Berry phase involved in the interference between 2 tunnelling
paths between the spin states A and B. From
Ref.~\cite{Wernsdorfer1999}}
\end{center}
\end{figure}

In closing these lecture notes, I wish to briefly mention some
further developments and applications of the concept of Berry
phase in magnetism.

In the previous section, we mentioned that the Berry phase can
give rise to interference phenomena for interfering paths in real
space. There may be also interferences associated with different
paths in spin space as well; this plays an important role in the
theory of tunnelling of magnetization in large spin molecular
magnets \cite{Loss1992b, vanDelft1992, Garg1993}. The situation is
sketched in Fig.~18. Depending on the value of the spin, and on
the solid angle between the 2 tunnelling trajectories from state A
to state B, interference due to the Berry phase take place; in
special cases, the interferences are destructive and tunnelling
becomes forbidden. This gives rise to very spectacular parity
effect that have been observed experimentally
\cite{Wernsdorfer1999, Wernsdorfer2002, delBarco2003}. For a
detailed review of this topic, see Ref.~\cite{Gateschi2003}.

The Berry phase plays a ubiquitous role in quantum mechanical
problems where one wants to treat the dynamics of some ``slow''
degrees of freedom, after having ``integrated out'' the ``fast''
degrees of freedom. An application of this concept in magnetism
concerns the adiabatic spin-wave dynamics of itinerant magnets.
Here the fast degrees of freedom are the electron degrees of
freedom giving rise to charge fluctuations and longitudinal spin
fluctuations, whereas the slow degrees of freedom are the
transverse spin fluctuations, i.e., the long wavelength magnons.
This was pioneered by Wen and Zee \cite{Wen1988}, and later on
further developed by Niu \emph{et al.} \cite{Niu1998, Niu1999}.
They obtained an equation of motion for the spins which is
controlled by the Berry phase; in the case of localized systems,
this reduces to the Landau-Lifshitz equation, but contains
non-local contributions in the case of strongly delocalized
systems.

For a spin $S$ coupled to a slowly moving magnetic field, we have
seen that the Berry phase is given by the solid angle described by
the field. For the case of a spin $S=1/2$, this situation
constitutes the most general case, however, for larger spins
$S\geq 1$, the most general Hamiltonian may contain more further
contributions, such as anisotropy terms, so that the parameter
space is much larger and richer than for a spin $1/2$. It is
therefore of interest to investigate the Berry phase in this more
general context. Recently, this has been investigating, by
considering more specifically the Berry phase associated with
global rotations of anisotropic spin systems \cite{Bruno2004b}.
This study reveals that beside the familiar solid angle term,
these is also a topological term, related to a winding number
giving the number of rotations of the systems around its
magnetization axis. This is relevant to any spin system of spin
$S\geq 1$, in particular to magnons ($S=1$), holes in
semiconductors ($S=3/2$), etc. A general theory of the Berry phase
of magnons is in preparation \cite{Dugaev2005}. Interesting
spin-wave interference phenomena have recently be obtained by
Hertel \emph{et al.} from micromagnetic simulations
\cite{Hertel2004}.

\newpage


\end{document}